\documentclass[a4paper,fleqn,usenatbib]{mnras}
\usepackage{epsfig}
\usepackage{amsfonts}
\usepackage{amsmath}
\usepackage{mathrsfs}
\usepackage{amssymb}

\newcommand{\mlya}{${\rm Ly\alpha}$}
\newcommand{\mlyb}{${\rm Ly\beta}$}

\usepackage{color}
\def\gdb#1{{\color{black}{#1}\color{black}}}

\def\review#1{{{#1}}}

\def\rereview#1{{{#1}}}

\title[The Thermal State of the Low-Density IGM]{Exploring the Thermal State of the Low-Density Intergalactic Medium at $z=3$ with an Ultra-High 
Signal-to-Noise QSO Spectrum}
\author[A.Rorai et al.]{A. Rorai$^{1,2}$\thanks{E-mail:arorai@ast.cam.ac.uk}, 
G.D. Becker$^{3}$, 
M.G. Haehnelt$^{1,2}$, 
R.F. Carswell$^{2}$, J.S. Bolton$^{4}$,
 S. Cristiani$^{5}$, 
\newauthor V. D'Odorico$^{5}$,
G. Cupani$^{5}$, P. Barai$^{6}$,
F. Calura$^{7}$,T.-S.Kim$^{5}$ E. Pomante$^{5}$,
\newauthor E. Tescari$^{8,9}$, M. Viel$^{5}$
\\
$^{1}$Kavli Institute for Cosmology and Institute of Astronomy, Madingley Road, Cambridge CB3 0HA, United Kingdom\\
$^{2}$Institute of Astronomy, Madingley Road, Cambridge CB3 0HA, United Kingdom\\
$^{3}$Department of Physics \& Astronomy, University of California, Riverside, 900 University Avenue, Riverside, CA,92521, USA\\
$^{4}$School of Physics and Astronomy, University of Nottingham, University Park, Nottingham, NG7 2RD, United Kingdom\\
$^{5}$INAF - Osservatorio Astronomico di Trieste, Via G.B. Tiepolo 11, I-34131 Trieste, Italy\\
$^{6}$Scuola Normale Superiore, Piazza dei Cavalieri 7, 56126 Pisa, Italy\\
$^{7}$INAF–Osservatorio Astronomico di Bologna, via Ranzani 1, I-40127 Bologna, Italy\\
$^{8}$School of Physics, The University of Melbourne, Parkville, VIC 3010, Australia\\
$^{9}$ARC Centre of Excellence for All-Sky Astrophysics (CAASTRO)}

\date{Accepted XXX. Received YYY; in original form ZZZ}

\pubyear{2016}

\begin{document}
\label{firstpage}
\pagerange{\pageref{firstpage}--\pageref{lastpage}}
\maketitle

\begin{abstract}

At low densities the standard ionisation history of the intergalactic medium (IGM) predicts a decreasing temperature of the IGM with decreasing density once hydrogen (and helium) reionisation is complete.  Heating the high-redshift, low-density IGM above the temperature expected from photo-heating is difficult, and previous claims of high/rising  temperatures in \gdb{low density} regions of the Universe based on the probability density function (PDF) of the  opacity in  \mlya\ forest data at $2<z<4$  have been met with  considerable scepticism, \gdb{particularly since they appear to be in tension with other constraints on the temperature-density relation (TDR).}  We utilise here an ultra-high signal-to-noise spectrum  of the QSO HE0940-1050 and a novel technique to study the low opacity part of the PDF. We show that there is indeed evidence 
\review{(at 90\% confidence level)} that  a significant volume fraction of the under-dense regions at $z \sim 3$ has temperatures as high or higher than those  at densities comparable to the mean and above.  
We further demonstrate that this conclusion is nevertheless consistent  with measurements of a slope of the TDR in over-dense  regions that imply a decreasing  temperature with decreasing density, as expected if photo-heating of ionised hydrogen is the dominant  heating process. We briefly discuss implications of our findings for the need to invoke \gdb{either spatial temperature fluctuations, as expected during helium reionization}, or additional processes that heat a significant volume fraction of the low-density IGM. 

\end{abstract}

\section{Introduction}

\gdb{The thermal state of the IGM in the redshift range $2 < z < 5$ has received considerable attention over the past two decades.  In part, this is due to the sensitivity of IGM temperatures to photo-ionization heating of the low-density IGM during helium (and potentially hydrogen) reionisation  \citep[e.g.,][]{HuiGnedin97,GnedHui98,TheunsMoSchaye01,Theuns02a,HH03}.  More generally, IGM temperatures are a potentially powerful diagnostic of galaxy and AGN feedback processes that can have lasting impact on the physical conditions of the low-density IGM.}

\gdb{The primary method to constrain} the thermal state of the IGM \gdb{has been to compare properties of the \mlya\ forest observed in quasar spectra to}  mock absorption spectra  obtained  
from cosmological hydrodynamical simulations \citep{Haehnelt98,Schaye00,Bolton08,
VielBolton09,Lidz09,BeckerBolton2011,Garzilli2012,Rudie2012,Boera2014,Bolton14,
KGpdf15}. 
In the numerical  simulations  the relation between temperature and density at low densities 
is generally well described by a simple 
power law of  the form 
\begin{equation}\label{eq:powerlaw}
T=T_0 \Delta^{\gamma-1} ,
\end{equation}
where $T_0$ is the temperature at the mean density of the Universe, $\Delta$ 
is the density in units of the mean and 
$\gamma$ is the index of the relationship. 
As first shown by \cite{HuiGnedin97}, such a power-law relation  arises 
due to the balance of photo-heating 
and the aggregate effect of recombination and adiabatic cooling due to the expansion of the Universe.

\gdb{The slope of the temperature-density relation is a potentially powerful tracer of both hydrogen and helium reionization.}  When  atoms are reionized, \gdb{photon energies in excess of the ionization potential are converted into}
thermal kinetic energy of the gas. In the simplest picture of an
instantaneous \gdb{and uniform reionization, this photo-ionization heating would impart a uniform amount of energy per atom at all densities, and hence produce a uniform temperature-density relation, i.e. $\gamma=1$.  In reality this picture may be significantly complicated due to inhomogeneous reionization and radiative transfer effects, which are likely to produce a multi-valued temperature-density relation during and shortly after reionization \citep{Trac08,McQuinn09,Compostella2013}.  Following hydrogen reionization, the IGM is expected to evolve towards a relatively simple thermal state.}  High densities are subject to higher recombination rates than 
low densities, implying multiple heating events, while the cooling is driven
by the adiabatic expansion of the Universe at all densities (radiative
cooling is negligible at IGM densities). Analytical
calculations and hydrodynamical simulations agree  that this 
behaviour leads to an asymptotic value of $\gamma \approx 1.6$ 
\citep{HuiGnedin97}.  Matters are further  complicated by the reionization of HeII, \gdb{however}, which appears to happen significantly later than hydrogen reionization
\citep[e.g.][]{Furlanetto08,Worseck2011}. 
Helium reionization should flatten again the temperature relation, and is also expected to be spatially inhomogeneous 
\citep{Abel99,McQuinn09,MeiksinTittley12,Compostella2013,Puchwein2015}. 
Following helium reionization, the IGM should again evolve towards an asymptotic thermal state with $\gamma \approx 1.6$.

\gdb{The potential for the thermal evolution of the IGM to shed light on} 
the reionization history 
\gdb{has motivated} numerous studies aimed 
at measuring the parameters of  the temperature-density relationship $T_0$ and $\gamma$. 
\gdb{While much progress has been made,} the results of these  studies 
\gdb{have not been} fully consistent with each
other. In particular, \gdb{some analyses} of the probability distribution 
function (PDF) of the transmitted \mlya\ flux \citep{kbv+07} have suggested
that the temperature-density relationship 
could be `inverted' (i.e. $\gamma <1$) at $z=2-4$ \gdb{\citep{Bolton08,VielBolton09,Calura2012}}. 
\gdb{As noted by \cite{Bolton08}, heating the low-density IGM} beyond the temperatures
expected from photo-heating is \gdb{physically challenging within the scenario outlined above.}
\gdb{It has been suggested that temperature inhomogeneities during helium reionization could help to explain the discrepancies between observed and simulated PDF \citep{McQuinn09,McQuinn2011b}.  A more speculative possibility is that a non-standard temperature-density relation could result from additional} 
heating at low densities due to plasma instabilities 
in the IGM following the  absorption of TeV photons
emitted by blazars that induce  pair-production (\emph{blazar heating})  \gdb{\citep{Blazar1,Blazar2,Blazar3}}. 
The efficiency of this process is still under debate \citep{Sironi2014},
but blazar heating  predicts density-independent (volumetric)  heat injection which would effectively
lead to an increasing temperature at the lowest 
densities \citep{Puchwein12,Lamberts2015}.

A number of follow-up studies have investigated  \gdb{the extent to which} the inferred 
$\gamma<1$ \gdb{could be} due  to  systematic errors in the analysis of  the \mlya\ forest data. In particular,
concerns have  been raised about  the uncertainty with regard to the continuum placement 
in quasar spectra \citep{Lee2012} and about the estimation of the errors from 
bootstrapping of the  data samples consisting of   small chunks of absorption spectra \citep{Rollinde2013}.
Skepticism about the claims that the \mlya\ flux PDF  suggests a temperature density 
relation with $\gamma \le 1$ appeared to be vindicated by    
careful  new measurements of $\gamma$ based on the line-fitting method \citep{Rudie2012,Bolton14}, which 
yield values in the more  'conventional' range $\gamma\sim 1.5-1.6$.
The line-fitting  technique is based on the decomposition of the \mlya\ forest into 
individual  Voigt profiles characterised by  an HI column
density  $N_{HI}$ and  Doppler  parameter $b$. 
Numerical simulations suggest  that the  distribution of line parameters in the 
plane defined by $N_{HI}$ and $b$ and in particular the low-b cut-off are  tightly related to 
the parameters characterising the temperature-density relation \citep{Schaye1999,Ricotti00,Bolton14}.

\begin{figure*}
\includegraphics[width=\textwidth]{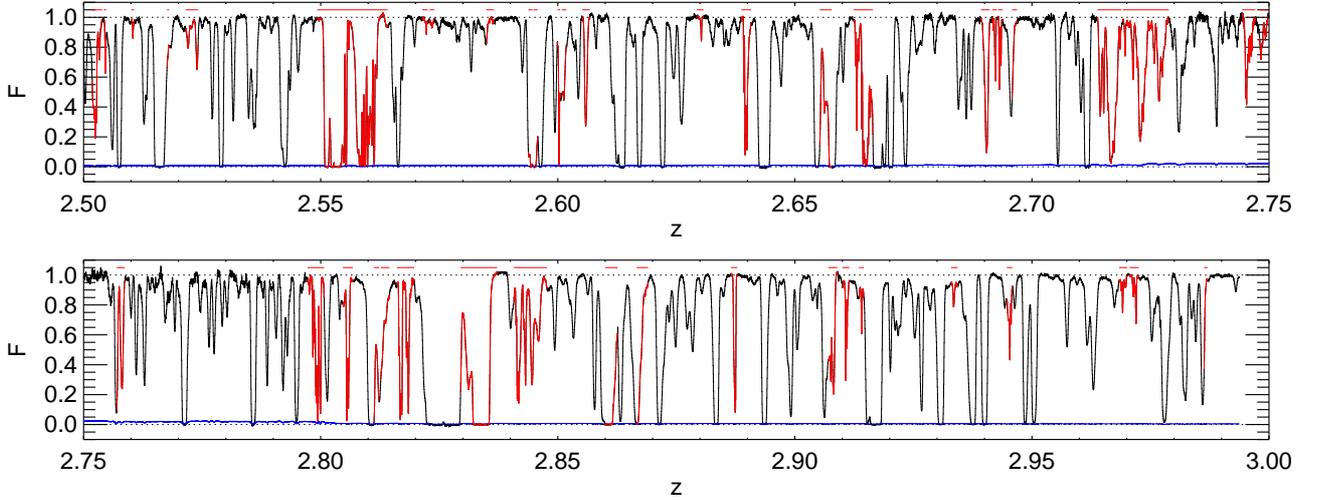}
\caption{\label{fig:deep_spectrum} The transmitted flux relative to the continuum level
in the \mlya\ forest of \review{HE0940-1050 ($z=3.0932$),} as a function of the 
redshift of \mlya\ absorption. 
In this work we analyse the region between $z=2.5$ and $z=3.0$,
which correspond to rest-frame wavelength of 1040 \AA\ and 1190 
\AA , respectively. 
In red we highlight the identified metal lines, which are 
masked within the 
region where the fitted absorption is smaller than 0.01 (in units
of the continuum, \review{these regions are marked by horizontal red lines above the spectrum).}
The blue line marks the noise level per pixel estimated by the 
\review{analysis}, while the black dotted lines trace the continuum 
($F=1$) and the zero-flux level.}
\end{figure*}

Other  methods to characterise the thermal state of the IGM that have been employed adopt the flux power spectrum \citep{McDonald2000,Zald01,Croft2002,Kim04} or closely-related statistics, like a decomposition into wavelets \citep{Lidz09,Garzilli2012} or the ``curvature'' \citep{BeckerBolton2011,Boera2014,Boera2016} of the observed flux. 
The common idea behind these analyses is to characterise the thermal broadening of \mlya\
lines by quantifying  the ``smoothness'' of the \mlya\ forest.
\gdb{Curvature methods have placed strong constraints on the} temperature of the IGM at a ``characteristic'' (over-)density,  $\bar{\Delta}(z)$, \gdb{which evolves with redshift; however, since $T(\bar{\Delta})$ is degenerate in combinations of $T_0$ and $\gamma$, the constraints on the shape of the temperature-density relation from these studies have so far been limited \citep[but see][]{Padmanabhan2015,Boera2016}.}

\gdb{A key challenge in reconciling discrepant constraints on $\gamma$ is to identify the physical properties of the IGM to which these statistics are most sensitive.  For example,} we  will see later that the results \gdb{from line fitting}
constrain the temperature density  relation  only for a rather limited 
range of densities. 
Matters are further complicated due to the fact that the thermal part of 
the broadening of \mlya\ forest absorbers is not only 
sensitive to the instantaneous temperature, but to the previous thermal 
history of the IGM. This is  due to the effect of 
thermal pressure  on the distribution of the gas, 
also commonly referred to as pressure smoothing 
\citep{Peeples09a,Peeples09b,Rorai13,Garzilli2015,Kulkarni2015}.

\gdb{In this study we will attempt to resolve the apparent tension between flux PDF and line-fitting constraints on $\gamma$ based on two factors: first, that the temperature-density relation at $z \sim 2-3$ may be more complicated than is often assumed, and second, that the two techniques sample the temperatures in different density regimes.  To illustrate this,} we take advantage of a \gdb{recently acquired} ultra-high S/N spectrum of the bright QSO 
HE0940-1050 ($z = 3.0932$).
The high S/N allows us to push our measurements to lower (over)-densities than previous studies 
and, more importantly, allows us an unprecedented control over the systematic effects due to possible 
errors in the continuum placement. To take full advantage of the high-quality spectrum we develop a novel 
analysis method that focuses on the high transmission part of the flux PDF and reduces the impact 
of continuum placement errors by re-normalizing the PDF to  a flux level close to the probability peak
(the 95th flux percentile).  We further use a suite of state-of-the-art hydrodynamical simulations for which we 
implement a wide range of temperature density relations in post-processing
to put quantitative constraints on a range of different parametrizations of the thermal 
state of the IGM with Markov chain Monte Carlo (MCMC) techniques.

The  paper is structured as follows. We  first describe the   data in 
\S~\ref{sec:data}.   We then illustrate in \S~\ref{sec:simulations} 
the simulations we use to build our IGM models and describe  
our different parametrizations of the thermal state of the IGM. Section \S~\ref{sec:method} 
presents the details of our analysis method, in particular our 
\gdb{approach to} dealing with continuum placement uncertainty, the optical depth 
renormalization, and the estimation of the PDF covariance matrix. Results are presented in \S~\ref{sec:results} and compared
with previous IGM studies in \S~\ref{sec:technique_comparison}. We
discuss the relevance of systematics effect and the  interpretation
of our results in \S~\ref{sec:discussion} and  summarize our main findings
in \S~\ref{sec:conclusions}.

\section{Data}\label{sec:data}

We analyse the spectrum of  HE0940-1050 ($z = 3.0932$) observed with
UVES \citep{UVES} at the VLT at a resolution of $R\simeq 45000$. 
This object was chosen as the
brightest target in the UVES Large Program \citep{Bergeron2004}. 
Our final spectrum (herein referred to as the ``DEEP’’ spectrum) was created by combining data from the ESO archive and data from a dedicated program of 43 hours
of observation carried out between December 2013 and March 2014.
The total integration time amounts to 64.4 hours. 
This observing time was estimated to be necessary to achieve a sufficient 
S/N in the \mlya\ forest and to allow the identification of weak
CIV and OVI metal lines (see D'Odorico et al., submitted).
The S/N in the  forest of the reduced spectrum is, on average, 280 
per resolution element.

The data has been reduced using the most recent version of the UVES 
pipeline \citep{Ballester2000} and a software developed by 
\cite{Cupani2015a,Cupani2015b} and binned in regular pixels of 2.5 km s$^{-1}$.
Special care was devoted to minimise the correlation between adjacent pixels of the spectra, which may cause an underestimation of the flux density error \citep{Bonifacio05}. Multiple wavelength re-binning of the data was avoided by using the non-merged order spectra (instead of the merged full-range spectra) when combining exposures taken at different epochs.

The final spectrum covers the wavelength range $\lambda \in 
[3050,7020]$ \AA\ in the vacuum-heliocentric reference system.  
We will focus here on the \mlya\ forest in the rest-frame 
wavelength range $\lambda \in [1040,1190]$ \AA\ , which for HE0940-1050
corresponds to the redshift interval  $z \in [2.5,3.0]$.
These limits are set in order to exclude the regions in proximity of the
\mlya\ and \mlyb\ lines. The continuum-normalized flux in
the relevant range is shown in figure \ref{fig:deep_spectrum}.

\section{Simulations}\label{sec:simulations}
\begin{center}
\begin{table}
\centering
\begin{tabular}{ccc}
\hline\hline\\
Model        &  Parameter & Description  \\
\hline\\
Standard   & $T_0$ & Temp. at $\Delta=1$  \\
\, & $\gamma$ & Index of the TDR \\
\,\\
Broken & $\Delta_b$ & Break density \\
\, & $T_b$ & Temp. at $\Delta=\Delta_b$ \\
\, & $\gamma_o$ & Index of the TDR at  $\Delta>\Delta_b$ \\
\, & $\gamma_u$ & Index of the TDR at  $\Delta<\Delta_b$ \\
\,\\
\rereview{Discontinuous} & $\Delta_b$ & Discontinuity density \\
\, & $T_+$ & Temp. at  $\Delta>\Delta_b$ \\
\, & $T_-$ & Temp. at  $\Delta>\Delta_b$ \\
\,\\
Fluctuations & $Q$ & Hot gas filling factor \\
\, & $T_h$ & Temp. at $\Delta=1$ in hot regions \\
\, & $T_c$ & Temp. at $\Delta=1$ in cold regions \\
\, & $\gamma_h$ & Index of the TDR in hot regions \\
\,\\
\emph{All} & $\xi$ & Smoothing parameter \\
\, & $\bar{F}$  &  Mean flux \\
\hline\hline
\end{tabular}
\caption{\review{Summary of the IGM parameters describing the 
different thermal models considered in the analysis.}}
\label{tab:sim}
\end{table}
\end{center}

\begin{figure}
\includegraphics[width=\columnwidth]{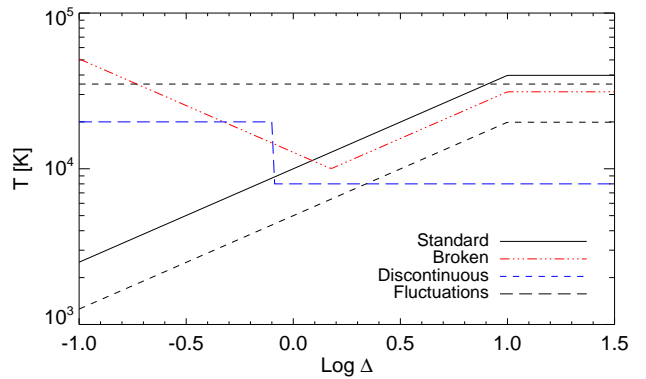}
\caption{\label{fig:models_summary} A summary of all the thermal models
we consider in our analysis. We first study the standard power-law
relationship between temperature and density (black solid) with parameters
$T_0$ and $\gamma$. We then generalize it by adding a break point at
some density $\Delta_b$. Such model, shown in red (dotted-dashed line),
would have a power-law index $\gamma_u$ in \review{underdense regions} 
and a different $\gamma_o$ in overdensities ($\Delta > \Delta_b$).
We also examine the case where the IGM has only two temperatures, below
and above some density threshold $\Delta_t$. 
In this plot, this translates into a step-function, 
depicted in blue (long-dashed line). Last, we implement a model where the IGM is divided
into hot isothermal regions and cold regions with $\gamma=1.6$. The 
division is done independent of density, and the two 
components are represented by the black dashed lines in the figure.}
\end{figure}

In order to predict the observed statistical properties of the 
\mlya\ forest, we used simulated spectra from the set of 
hydrodynamical simulations described in \cite[][\gdb{hereafter} B11]{BeckerBolton2011}.  The simulations were run using
the parallel Tree-smoothed particle hydrodynamics (SPH) code
GADGET-3, which is an updated version of the publicly available code
GADGET-2 \citep{Gadget2}.
The fiducial simulation volume is a 10 Mpc$/h$ periodic box
containing $2 \times 512^3$ gas and dark matter particles. This resolution
is chosen specifically to resolve the Ly$\alpha$ forest at high redshift
\citep{Bolton2009}. The simulations were all started at 
$z = 99$, with initial conditions generated using the transfer function of \citep{Eisenstein99}. 
The cosmological parameters are $\Omega_m=0.26,\Omega_{\lambda}=0.74,\Omega_{b}h^2
=0.023,h=0.72,\sigma_8 = 0.80, n_s = 0.96,$
consistent with constraints of the cosmic microwave background from 
WMAP9 \citep{Reichardt2009,Jarosik2011}. The IGM is assumed
to be of primordial composition with a helium fraction by mass
of $Y=0.24$ \citep{Olive2004}. The gravitational softening
length was set to 1/30th of the mean linear interparticle spacing and
star formation was included using a simplified prescription which
converts all gas particles with overdensity $\Delta = \rho/\bar{\rho}>10^3$ and
temperature $T < 10^5$ K into collisionless stars. In this work we will only
use the outputs at $z=2.735$ .

\begin{figure*}
\includegraphics[width=\textwidth]{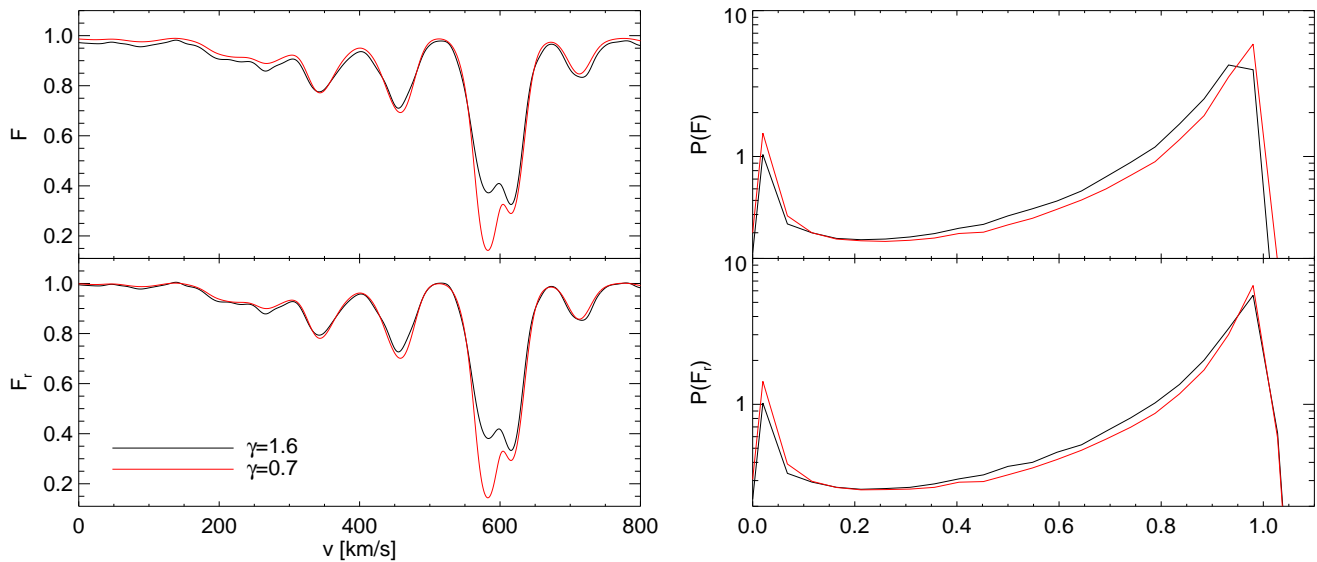}
\caption{\label{fig:continuum_regulation}
\review{ \emph{Left column:} the continuum regulation procedure applied to a 
simulated spectrum calculated assuming
either $\gamma=1.6$ (black) or
$\gamma=0.7$ (red). In the upper panel we plot the flux level
in the two models, in the lower panel we show the same spectra
after the percentile regulation.
The problem of the continuum placement  can be
understood by carefully looking at the difference between the red
and black lines at high fluxes: without regulation 
(upper panel) the black model ($\gamma=1.6$) is 
systematically \emph{lower} than the red one, which could easily lead
to an underestimation of the continuum level.  
With the regulation  (lower panel) we impose an alignment
in correspondence of the 95th percentile of the flux distribution, 
regardless of the initial placement of the continuum.\label{fig:regulated_pdf}
\emph{Right column:} the PDF of the flux $F$ (upper panel) and of the 
continuum regulated flux $F_r$ (lower panel), for the same 
two IGM models. By imposing
the percentile regulation we artificially align
the peaks of the distributions, but sensitivity
to the thermal state is retained by the overall shape. The advantage 
of this method is that the regulated PDF is less sensitive to continuum
placement errors. }}
\end{figure*}

The gas in the simulations is assumed to be optically thin and
in ionization equilibrium with a spatially uniform ultraviolet background
(UVB). The UVB corresponds to the galaxies and quasars
emission model of \cite{HM01} (\gdb{hereafter} HM01). Hydrogen is reionized
at $z=9$ and gas with $ \Delta \lesssim 10$ subsequently follows a tight
power-law temperature-density relation, $T=T_0 \Delta^{\gamma-1}$, where $T_0$
is the temperature of the IGM at mean density \citep{HuiGnedin97,
Valageas2002}. As in B11 , the photo-heating
rates from HM01 are rescaled by different constant 
factors, in order to explore
a variety of thermal histories.  
Here we assume the photo-heating rates $\epsilon_i=\xi \epsilon_i^{HM01}$, where 
$\epsilon_i^{HM01}$ are the HM01 photo-heating rates for species
$i=$[HI,HeI,HeII]  and $\xi$ is a free parameter. 
Note that, differently than in B11, we do not consider models 
where the heating rates are density-dependent.  
In fact, we vary $\xi$ with the only purpose of varying the 
degree of pressure smoothing in the IGM, while the TDR
is imposed in post-processing. In practice, we only 
use the hydrodynamical simulation to obtain realistic density and
velocity fields. For this reason, we will often refer to $\xi$ as
the 'smoothing parameter'.
We then impose a specific temperature-density relationship on top of 
the density distribution, instead of assuming the temperature 
calculated in the original hydrodynamical simulation. We opt for
this strategy in order to explore a wide range of  
parametrizations   of the thermal state of the IGM, at the price of 
reducing the phase diagram of the gas to a deterministic relation
between $T$ and $\rho$. \gdb{In practice, we find that the native TDR 
and the matching best-fitting power laws produce nearly identical flux PDFs.}

Finally we calculate the optical depth to \mlya\ photons for a set of 
1024 synthetic spectra, assuming that the gas is optically thin, 
taking into account 
peculiar motions and thermal broadening.  \gdb{For our fiducial spectra we scale the UV background
photoionization rate $\Gamma$ in order to match the observed mean flux 
of the forest at the central redshift of the DEEP spectrum \citep[$\bar{F}_{\rm obs}(z=2.75)=0.7371$, ][]{Becker2013}.  As described below, however, the mean flux is generally left as a free parameter when fitting the data.}

We stress that in this scheme the pressure smoothing and the temperature 
are set independently.
\gdb{While not entirely physical, this allows us to separate the impact on the Ly$\alpha$ forest from instantaneous temperature, which}
depends mostly on
the heating at the current redshift, \gdb{from pressure} smoothing, \gdb{which} is a result of the 
integrated interplay between pressure and gravity across the whole thermal history
\citep{GnedHui98}.

\subsection{Parametrizations of the Thermal State of the IGM}\label{sec:parametrizations}

In this section we summarize the thermal models \review{and parameters considered in our 
analysis. 
The first parameter is the heating rescaling factor, $\xi$,} 
\gdb{used on the hydrodynamical simulations, which determines the amount of pressure smoothing.  We also include the mean transmitted flux,} $\bar{F}$, in the parameter set; however, rather than 
allow it to vary freely, we impose a Gaussian prior based on the recent
measurement of \cite{Becker2013}: 
\begin{equation}\label{eq:prior}
p(\bar{F})\propto \exp\left[-\frac{(\bar{F}-\bar{F}_{obs})^2}{2\sigma_{\bar{F}}^2}\right],
\end{equation}
where $\bar{F}_{obs}=0.7371$ and $\sigma_{\bar{F}}=0.01$ (which is
slightly more conservative than the estimated value). 

\gdb{In order to test whether our findings reproduce previous results in the literature,} we first explore the standard power-law TDR parametrized by $T_0$ and $\gamma$ \gdb{(equation~\ref{eq:powerlaw}).}
The full parameter set
describing this model is therefore $\{T_0,\gamma,\xi,\bar{F} \}$.

\begin{figure*}
\includegraphics[width=\textwidth]{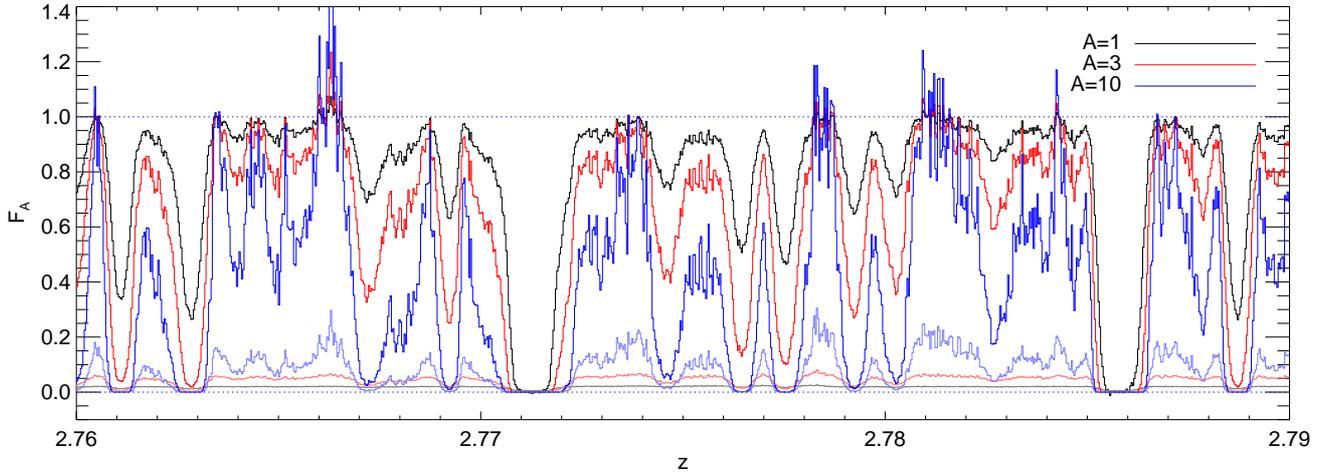}
\caption{\label{fig:spectrum_rescaling}
The effect of rescaling the optical depth for a selected 
chunk of the \mlya\ forest of the DEEP spectrum. The plot shows the
transformed flux $F_A$ after multiplying $\tau$ by a factor $A=1$ (black, 
original spectrum), 3 (red) and 10 (blue).  The spectrum is 
continuum-regulated before applying the
transformation, so that 5\% of the pixels have flux above 1, explaining
the occasional spikes in the transformed flux. The propagated instrumental
errors after the transformation are shown as lighter lines at the 
bottom, matched by color to the transformed fluxes. The motivation for 
applying this transformation is that it amplifies fluctuations near the
continuum \review{as A increases}, while the most absorbed regions, corresponding to 
high densities, become saturated. This mimics what happens in a 
denser universe or with a lower amplitude UV background, resembling the \mlya\ 
forest at higher $z$. In this way we are able to increase our 
sensitivity to the low density regions of the IGM, which are difficult
to probe otherwise.}
\end{figure*}

\gdb{Next} we consider the possibility that underdense regions
 and overdensities follow  
different power-law relationships.  \gdb{This model is partly} inspired by blazar-heating models, \gdb{but is mainly designed to provide a simple extension to the simplest power-law model.} 
The relation between temperature and density \gdb{is} described
by the expressions
\begin{equation}
T(\Delta)=T_b \times \begin{cases}
\left(\frac{\Delta}{\Delta_b}\right)^{\gamma_{\rm u}-1} & {\rm if\ } \Delta < \Delta_b , \\
\left(\frac{\Delta}{\Delta_b}\right)^{\gamma_{\rm o}-1} & {\rm if\ } \Delta_b \leq \Delta < 10 ,\\
{\rm constant} & {\rm if\ } \Delta > 10 .
\end{cases}
\end{equation}
Here, $\Delta_b$ is the 'break' density where the transitions between the
two regimes occurs. 
We define a threshold for $\Delta > 10$ to avoid
unrealistically high temperatures in the high-density regions.
This however has little practical effect, as such densities represent
a tiny fraction of the \mlya\ forest pixels.
The parameters 
$\gamma_o$ and $\gamma_u$ define the slope above and below the break
density.
This
parametrization is thus defined by $\{T_b,\Delta_b,\gamma_u,\gamma_o,
\xi,\bar{F}\}$. Note that the standard temperature-density relationship
represents a special case of this set of models.

We also calculate a grid of models where the TDR 
is described by a simple step function: 
\begin{equation}\label{eq:step_function}
T(\Delta)= \begin{cases}
T_{-} & {\rm if} \Delta < \Delta_b , \\
T_{+} & {\rm if} \Delta \geq \Delta_b . 
\end{cases}
\end{equation}
In  this  model the temperature distribution of the  IGM is assumed to be bimodal  with only two temperatures,
$T_-$ in underdense regions and $T_+$ in overdensities. 
Each model is fully defined
by the quintuplet $\{T_+,T_-,\Delta_b,\xi,\bar{F}\}$. 
The motivation behind this simple model is to understand the effect of a 
density-dependent temperature contrast in its simplest form, rather than
assuming some special functional form for the relationship between 
temperature and density. 

Lastly, we consider a set of models \gdb{including multiple temperatures at a given overdensity.  These are meant to mimic the temperature fluctuations expected to be present during helium reionization; however, we do not attempt to capture the full complexity of realistic fluctuations.  Instead, we employ a simple model wherein we consider}
two regions independently of the density. 
\gdb{The ``hot'' regions have a} temperature-density relationship  
defined by $T=T_h\Delta^{\gamma_h -1}$, where $T_h$ is the temperature
at mean density, ranging between 15000 K and 35000 K, and 
$\gamma_h \in [0.4,1.3]$ is the index. The ``cold'' regions 
are characterised by $T=T_c\Delta^{\gamma_c-1}$, with $T_c \in [5000,15000]$ K
and $\gamma_c=1.6$, as expected for the IGM long after a reionization
event. The fraction of space occupied by the hot region \gdb{is set by a}
filling factor, $Q$.  All together, the fluctuation models are defined 
by the parameter set $\{T_h,\gamma_h,T_c,Q,\xi,\bar{F}\}$.

\rereview{Table~\ref{tab:sim} recapitulates the four kinds of 
models analysed in this work and lists the parameters that
characterize them. 
A visual summary of all the thermal models 
is given in figure \ref{fig:models_summary}. }

\section{Calculation of the Regulated Flux Probability Distribution}\label{sec:method}

In order to exploit the \gdb{exceptionally high signal-to-noise ratio} of \gdb{our UVES spectrum of HE0940-1050}, we 
apply several modification to the standard flux PDF, which are described 
in the following sections. 

\begin{figure*}
\includegraphics*[width=\textwidth]{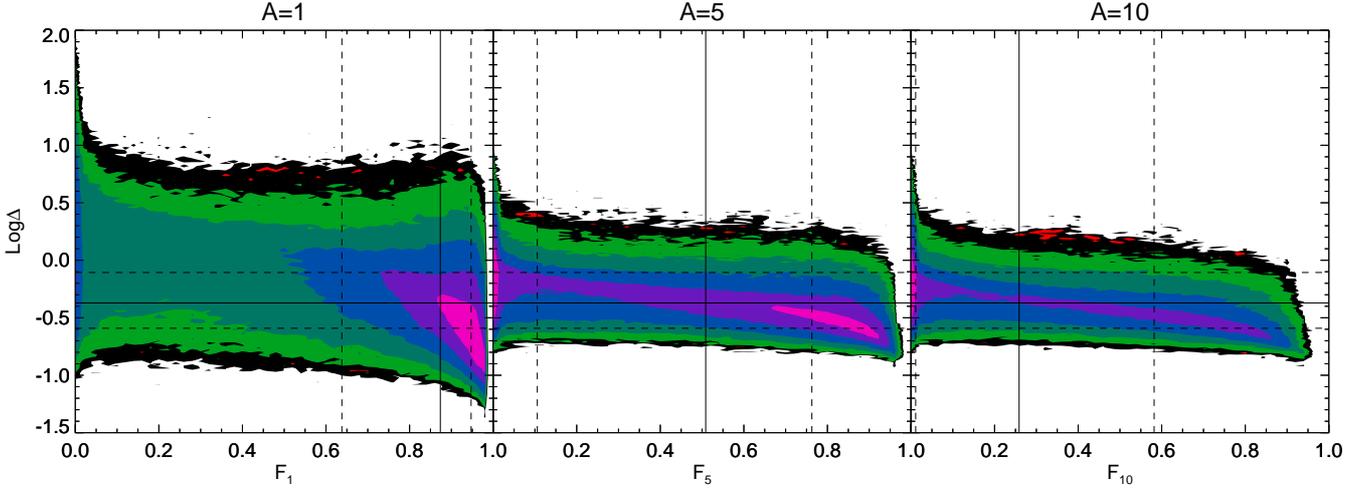}
\caption{\label{fig:flux_den_space}
Contour plot of the 2-d distribution of pixels from a simulation in
flux-density space. The density pixels, differently than the flux,
are calculated in real space. Each color step represents a change in 
pixel frequency of half a dex (arbitrary units). 
The simulation has a pressure smoothing parameter of $\xi=0.8$, and the flux
used in the left panel is directly taken from the output of the 
hydrodynamical simulations (i.e. it is not calculated by imposing a 
temperature-density relationship in post-processing). The thermal
parameters fitted to the temperature-density distribution of this
simulation are $T_0=9600$ K and $\gamma=1.54$ at $z=3$.
In the central and right panels we have rescaled the optical 
depth by a factor 5 and 10, respectively. 
The  vertical (horizontal) black solid line marks 
the \emph{median} of the flux (density) distribution, while the 
dashed lines are drawn in correspondence of the 25th and 75th 
percentiles.  
In the right panel, we note that more than 75\% of the pixels 
corresponds to densities
below the mean, and more than 75\% have a transmitted flux 
greater than 0.6, occupying the part of the PDF around the peak.
Rescaling the optical depth guarantees a more regular distribution
of pixels along the flux dimension. Furthermore, with a 
rescaling factor of 5 or 10,  the density distribution is 
centred below the mean density  \emph{for any value of the flux}.
This transformation is thus well suited to study the under-dense regions.}
\end{figure*}

\subsection{Noise and Resolution Modelling}\label{sec:forward_modelling}

Instrumental effects are taken into account by including them
in the 
synthetic spectra. 
The \gdb{finite} resolution is mimicked by convolving the 
predicted \mlya\ flux with a Gaussian kernel
of width FWHM$\approx 7.2$ km/s, \gdb{appropriate}
for a slit aperture of 1''. 
The simulated spectra are then rebinned onto a regular pixel grid with spacing $\Delta v= 2.5$ km/s, 
as in the DEEP spectrum.

\gdb{Noise is added to the simulated skewers by the following procedure:}
\begin{enumerate}
	\item We group the pixels of the Deep Spectrum into 50 
	bins according to the transmitted \mlya\ flux.
	
	\item The instrumental errors of these pixels define 50
	different noise distributions, representing the possible value of
	the noise for a given \mlya\ flux. 
	
	\item For each simulated pixel, we randomly select a noise value 
	from one of the 50 subsets above, according to the flux.
\end{enumerate}

Note that this procedure does not capture the wavelength coherence 
of the noise amplitude  (namely, the magnitude of errors should be correlated
across adjacent pixels). This does not affect the PDF, however,
which retains no information about the spatial distribution of 
the data. 
A test of the importance of noise and resolution modelling is 
presented in appendix \ref{appendix:noise}.

\subsection{Continuum Uncertainty}\label{sec:continuum_regulation}

Continuum \gdb{placement} is known to be of critical importance in studying
the flux PDF of the \mlya\ forest \citep[e.g.,][]{Becker2007,Lee2012}. Misplacing the continuum level
produces a multiplicative shift in the flux values,
causing a 'compression' or a stretching of the PDF which may lead
to significant bias on the constraints on $\gamma$. 
This is of particular concern in the present work, 
as we aim at characterizing the absorption distribution at high flux levels. 

The problem of dealing with continuum uncertainty has been sometimes 
addressed by trying to estimate it based on the unabsorbed part of the
spectrum, on the red side of the \mlya\ emission lines.
Typically, such methods are built upon either a power-law extrapolation 
\citep[e.g.,][]{Songaila2004}, or a Principal Component analysis 
of a training set of low-redshift quasars  \citep{KGPCA2012}. 

\gdb{Here we adopt a complementary approach that reduces
the sensitivity of the flux PDF to the continuum fitting.} 
Assuming that the continuum 
uncertainty is well approximated by 
a multiplicative parameter,
we can eliminate it  by adopting a 
'standard' renormalization based on the spectrum properties. 
For example, one may choose to set the level $F=1$ at the maximum
transmitted flux. Of course, this option is not optimal because the 
maximum would depend on the noise properties. Alternatively, one can
\gdb{adjust the continuum so that the mean normalized flux} 
matches the observed value 
at the equivalent redshift \citep{KGPCA2012}.  
\gdb{This ignores}
variability of the mean flux from one line sight
to another, however, which can add significant noise to the measurement. 
We opt for the following solution. Taking \gdb{the initial fitted continuum as a starting point, we divide the spectrum into 10 Mpc/$h$ regions.  In each region, we find the} 
flux level, $F_{95}$, corresponding to the 
95$^{\rm th}$ percentile of the distribution. We then \review{define} the ``regulated'' 
flux \gdb{in each region} as $F_r=F/F_{95}$, \gdb{and compute the PDF of $F_r$}. 
The advantage of using $F_{95}$ is that the 
95-th percentile falls near the peak of the flux PDF 
for all the IGM models, and it is therefore less noisy than the mean,
which falls in a flux interval of low probability.

Realistically, one might expect that spectral features like emission 
lines would demand a more complex modelling of the continuum 
uncertainty.
Fortunately, however, 10 Mpc$/h$ (equivalent
to 1014 km/s \review{at $z\approx 2.75$}) is smaller than the typical width of emission 
lines in quasar spectra. For consistency, the same kind of regulation is
applied to the simulated spectra we use to compare the observation with. 
We perform a test of the effectiveness of the regulation procedure
using a set of mock spectra, which is presented in appendix
~\ref{sec:continuum_test_mocks}.
In figure \ref{fig:continuum_regulation} (left column) \review{we report  an example of a
simulated spectrum before (upper panel) and after (lower panel) 
the percentile regulation}, calculated assuming a thermal model with 
$\gamma=1.6$ (black) and $\gamma=0.7$ (red). Both models have 
$T_0=15000$ K. Regulation has the effect 
of aligning the spectra of the two models at high fluxes.

We stress that the statistic obtained in this way (which we will call 
the \emph{regulated} flux 
PDF) is different than the standard PDF analysed in
previous \review{works. The right column of Fig.~\ref{fig:regulated_pdf}
demonstrates that although the regulation aligns the position of the peak,
the overall shape retains information about the thermal state.}

\subsection{Optical Depth Rescaling}\label{sec:transformation}

\review{ \mlya\  absorption traces 
different density ranges at different redshifts.} The leading factor in
this evolution is the expansion of the Universe. As the overall density
decreases, higher overdensities are required to produce the same amount of 
absorption. 
This has been quantified in B11, in the context of the 
the  temperature  measurement based on the "curvature" method, 
by \gdb{identifying} a characteristic density, $\bar{\Delta}$, as 
a function of redshift, \gdb{at which}
the corresponding temperature is a 
bijective relation with the mean curvature of the \mlya\ forest 
The existence of $\bar{\Delta}$
has been broadly interpreted suggesting that the \mlya\ probes that particular
density. For the redshift range spanned by the DEEP spectrum, the 
value of the characteristic density calculated in B11 is $\bar{\Delta}=3.35$.

\gdb{We can partially modify the densities to which the \mlya\ 
forest is sensitive by implementing}
a convenient transformation of the 
flux field. 
With the aim of studying the low-density range of the IGM, 
we can artificially enhance the 
\mlya\ optical depth in each pixel of the spectrum by some arbitrary factor $A$
\begin{equation}
\tau_A=A\tau ,
\end{equation}
where $\tau$ is the observed optical depth. In term of the transmitted
flux, this translates into 
\begin{equation}
F_A=\exp(A \log|F|) =|F|^A ,
\end{equation}
where the absolute value is taken to make the \review{transformation} well-defined 
for negative values of the observed flux $F$. All pixels are in this way forced
to have a positive flux, however this will not affect the shape of the PDF
since \review{negative pixels are generally} contained in the 
 bin centred around $F=0$, which is excluded from the analysis (see below).

The effect of this transformation on a
section of the DEEP spectrum can be seen in figure \ref{fig:spectrum_rescaling}.
Compared to the original \mlya\ flux (black), the rescaled spectra with 
$A=3$ (red) and $A=10$ (blue) are suppressed at low
fluxes, while fluctuations close to the continuum are progressively 
amplified.  At high flux levels, the transformation amplifies both real fluctuations
and instrumental noise (shown as thin lines).

\gdb{It is therefore a distinct advantage to start with}
a spectrum of exceptional quality like the DEEP spectrum.
Figure \ref{fig:flux_den_space} shows the distribution of 
pixels from simulated spectra in the flux-density plane, 
with and without optical-depth rescaling ($A=1,5,10$ from left
to right). From this plot it
is possible to appreciate the density range probed by 
the forest at each flux value.  The vertical lines 
mark the 25\% (dashed), 50\% (solid) and 75\% (dashed)  percentiles
of the flux distribution, giving a different perspective 
on the effect of the optical depth rescaling. Note however that the
flux and the optical depth are defined in velocity space, while the
density is defined in real space, so the two quantities are related 
only in an approximate and statistical sense. 

\begin{figure}
\includegraphics[width=\columnwidth]{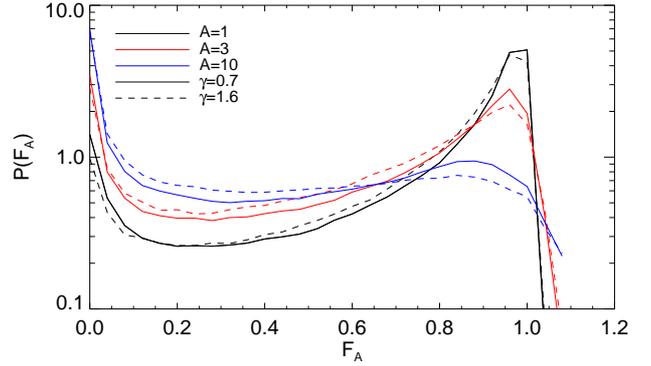}
\caption{\label{fig:pdf_rescaling}
The effect of the optical depth rescaling on the flux
PDF, for $A=1,3,10$ (black, red and blue, respectively). The
solid lines are calculated from an "inverted" model with 
$\gamma=0.7$ and $T_0=15000$ K, while the dashed
lines represent a model with the same $T_0$ but $\gamma_u=1.6$.
This figure illustrates the advantage of applying the rescaling
to the optical depth: as $A$ increases, the structure of the 
peak of the  PDF is better resolved, and the difference between 
the two models is more evident. In other words, the rescaling
corresponds to 
a convenient rebinning of the flux where the data are more 
evenly distributed over the bins, in particular near the 
continuum level. }
\end{figure}

\gdb{The impact of the transformation on the flux PDF is}
illustrated in Fig.~\ref{fig:pdf_rescaling}. A significant 
fraction of the pixels are contained in the peak of the \gdb{(untransformed)} PDF 
around $F\approx 0.9$. The structure of this peak is sensitive
to the temperature-density relationship at low densities. 
By applying the optical depth rescaling, the pixels at the 
peak are redistributed \gdb{over a larger} number of bins, as is clear
by comparing the red and the blue curves to the black one, which is 
obtained without rescaling. This is equivalent to change the binning 
of the standard PDF in a flux-dependent way, such that the sensitivity
at high fluxes is enhanced.  
In this work we will adopt $A=10$ for the data analysis and to obtain
the parameters constraints. We have preliminarily tested that
the results are not 
very sensitive to this choice as long as $A>5$.
In the following, we will refer to the 
PDF of the transformed flux $F_{10}$ as the \emph{transformed} PDF.
\gdb{We stress that this transformation is applied after the continuum regulation described in \S~\ref{sec:continuum_regulation}.} 
\review{The tabulated values of the PDF obtained in this way can be 
found in Table~\ref{tab:pdf} at the end of this manuscript.}


\subsection{Metal Line Contamination and Lyman Limit Systems}
\label{sec:contaminants}

\begin{figure*}
  \vskip -0.2in
     \centering{\epsfig{file=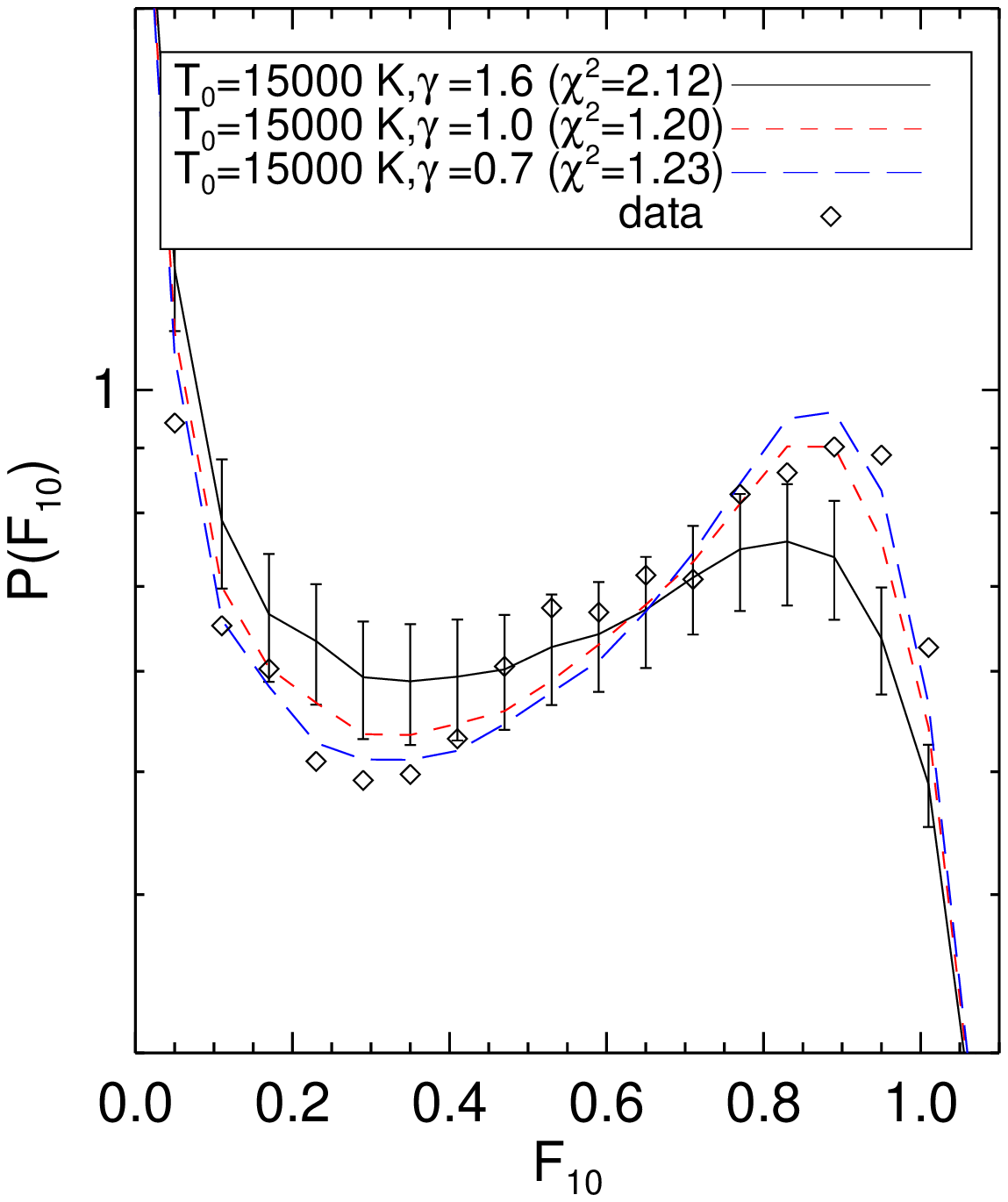, width=0.3\textwidth,clip}
     \epsfig{file=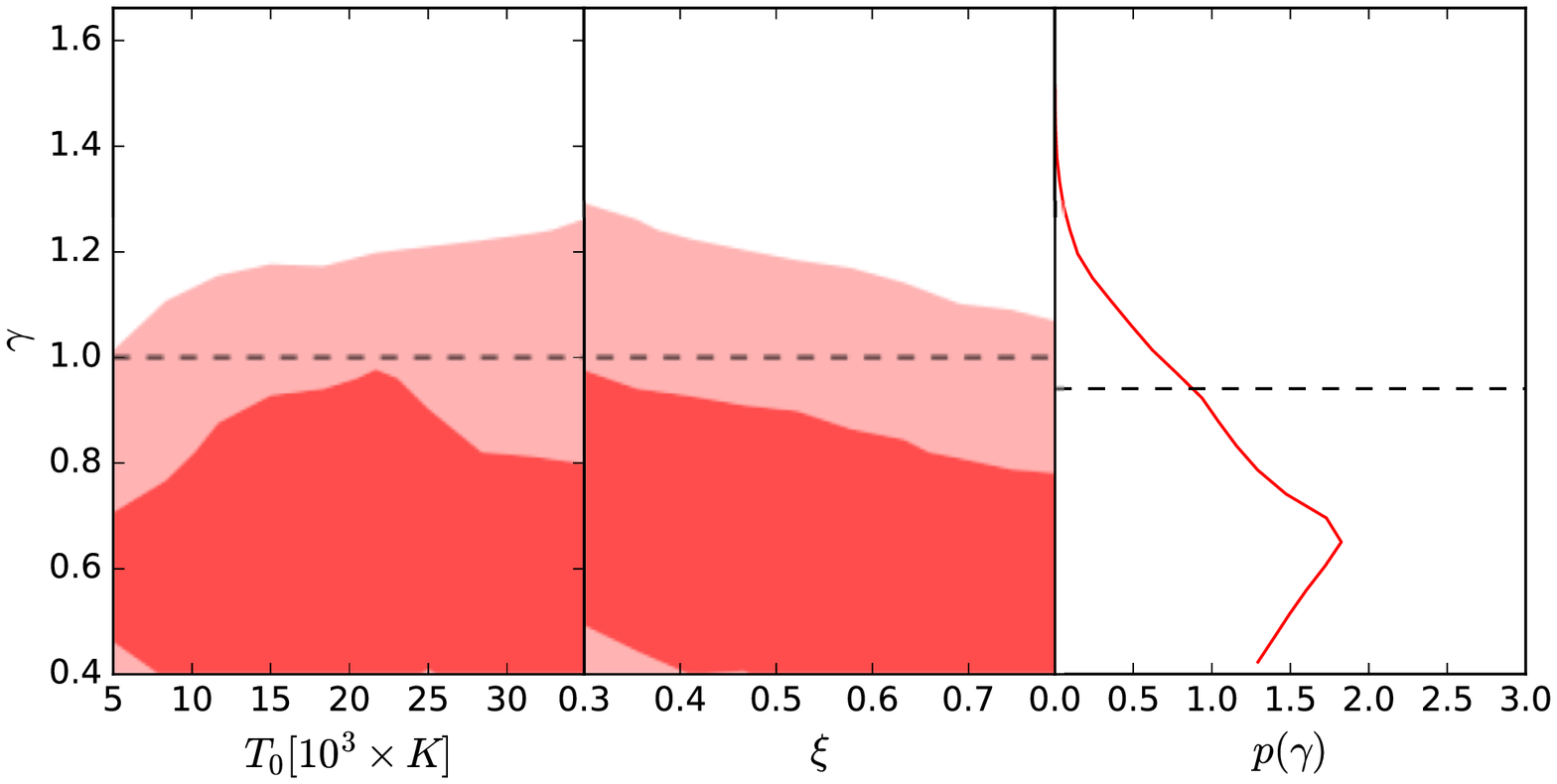
       , width=0.66\textwidth,clip}}
       
	\centering{}


\caption{\label{fig:t0gamma}
\review{\emph{Left panel:}} PDF of the regulated and transformed flux 
$F_{10}$ of the DEEP spectrum (black diamonds), compared with the prediction
of three different models of the thermal state of the IGM. The standard $\gamma=1.6$ model is 
shown in black, with errorbars estimated from the simulated spectra as 
explained in the text. The flat ($\gamma=1$, red dashed) and inverted 
($\gamma=0.7$, blue long-dashed) models provide a better fit to the observations,
confirming previous findings from the flux PDF. \review{To facilitate
the comparison between different models, the lowest-flux bin is not shown. In the legend we provide the values of  
the reduced $\chi^2$ for the three models, calculated taking into account the full covariance 
and assuming 12 degrees of freedom (16 independent bins and 4 free parameters). The lowest reduced-$\chi^2$ value achieved in our parameter grid 
is $\chi^2=0.92$.}
\label{fig:t0gammaconstraints} 
\review{\emph{Right panel:}} results of the MCMC analysis applying the likelihood in 
eqn.\ref{eq:likelihood} to the PDF of the transformed flux
of the DEEP spectrum (where the optical depth is rescaled by 
$A=10$). The plot shows the 1-$\sigma$ 
and 2-$\sigma$ confidence levels in the $T_0$-$\gamma$ plane (left
panel), in the $\xi$-$\gamma$ plane, as well as the marginalized 
posterior distribution for $\gamma$. The dashed horizontal line marks the 
level $\gamma=1$, as in an isothermal IGM. 
We assume a Gaussian prior for the mean flux with mean at 
$\langle F \rangle = 0.7371$ and standard deviation of 
$\sigma_{\langle F \rangle}=0.01$, based on previous measurements. 
With these assumptions, the PDF strongly favours an inverted
temperature-density relationship \review{if parametrized as a 
simple power law}, consistent with results in the 
literature \protect\citep{Bolton08,VielBolton09,Calura2012}.}
\end{figure*}

\gdb{Intervening metal lines will tend to modify the flux PDF by increasing the overall opacity
of the forest.  Rather than attempt to include metals in our simulated spectra, we take
advantage of the high quality of the DEEP spectrum by carefully identifying and masking
regions contaminated by metal absorption.  Our procedure is summarized as follows:}
\begin{itemize}
	\item We first identify all \gdb{ metal absorption systems using lines that fall redward of} 
	the \mlya\ 	emission line. 
	
	\item \gdb{We then identify all regions within the forest that are potentially contaminated by metal species associated with these absorbers.}
		
	
	\item Finally, \gdb{remaining} lines within the forest \gdb{that have conspicuously narrow Voigt profile fits (with Doppler parameter
	$b<8 $ km s$^{-1}$) are noted and identified where possible.} 
	
	\item Potential \mlya\ lines which do not present compatible higher order Lyman lines (i.e. they are too weak) are also classified, and subsequently identified, as metal lines.
	
\end{itemize}

At the end of this process we have a list of lines which we use to
define our masking.  \gdb{We perform Voigt profile fits on these lines, and} exclude from the analysis all regions where 
metal absorption amount to more than 1\% of the unabsorbed continuum level.  
This resulted in a cut of about $20 \%$ of the total \mlya\ forest, leaving
an effective path length of about $\Delta v_{\rm TOT}=31320$ km s$^{-1}$. 

The list of ions we consider includes: CII, CIV, SiII, SiIII, SiIV, AlII, AlIII, FeII, MgI, MgII, CaII.  

The remaining lines are fitted assuming that they are HI \mlya\  absorbers.
Among these 
there are two systems for which the HI column density is
estimated to be $N_{HI}>10^{17.2} $ cm$^{-2}$, at $z=2.861$ and $z=2.917$. 
Such systems are classified as Lyman limit systems (LLS) and 
are known to be self-shielding systems in which the optically
thin approximation does not hold \citep[e.g.][]{Fumagalli2011},
\review{so we mask them out from the spectrum. }

Because this process is subject to unavoidable uncertainties, we have carried out a test
\emph{a posteriori} where we explicitly show that our results are very
weakly sensitive to the removal of metal lines 
(see appendix \ref{appendix:contaminants}).

%

\subsection{Error Calculation and Likelihood Function} \label{sec:error_estimation}

A reliable estimate of the errors on the PDF is not achievable 
with a single spectrum, let alone the calculation of the full  
covariance matrix. Therefore we follow a complementary approach in
which uncertainties are estimated from the simulations.
We extract $n=1000$ mock samples of spectra with the 
same path length as the DEEP spectrum (31320 km s$^{-1}$). Flux
PDFs are calculated for these samples, \gdb{from} which we 
compute the covariance matrix as 
\begin{equation}
\Sigma_{i,j}=\frac{1}{n}\sum(p_i-\bar{p_i})(p_j-\bar{p_j}),
\end{equation} 
where $p_i$ is the PDF value in the $i-$th flux bin and the sum
is performed over the ensemble of 1000 mock samples. 
We further discuss the calculation and
the convergence of the covariance in \S~\ref{sec:discussion} and 
\review{appendix}~\ref{app:convergence}. In particular, we show that due to cosmic
variance we are underestimating variances by  $\sim 44\%$,
\gdb{which we correct by}
multiplying each element of $\Sigma_{i,j}$ by 1.44.  

The likelihood is then defined following the standard assumption
that the 18 flux bins are distributed as a multivariate Gaussian. 
Two degrees of freedom are removed by the normalization 
condition ($\Sigma p_i =1$) and by the percentile regulation, hence
we remove from the analysis the highest and the lowest of the flux
bins. 
Once this is done, the likelihood is defined as 
\begin{equation}\label{eq:likelihood}
L=\frac{1}{\sqrt{(2\pi)^k |\Sigma|}}\exp\left[-\frac{1}{2}(p_m-p_d)^T\Sigma^{-1}(p_m-p_d) \right]
\end{equation}
where $|\Sigma|$ denotes the determinant of the covariance matrix, $p_m$ the 
PDF array predicted by the model and $p_d$ the PDF measured from the data.
We stress that \gdb{both} the predicted PDF $p_m$ \gdb{and the} covariance matrix are model-dependent.

\subsection{MCMC analysis}

\begin{figure*}
\includegraphics[width=\textwidth]{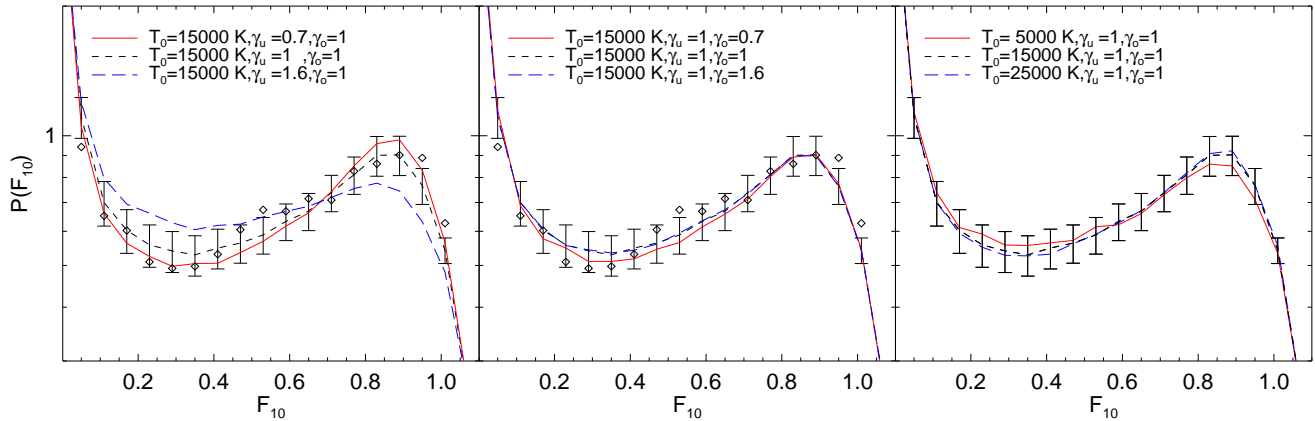}
\caption{\label{fig:broken_trho_pdf} 
Probability function of the regulated and transformed flux 
of the deep spectrum (black diamonds), compared with the prediction
of IGM models with a temperature-density relationship described by 
a broken power law. \emph{Left panel:} the isothermal model 
(black dashed with errorbars) is taken as a reference, from which we vary the index of the TDR
in \review{underdense regions} to $\gamma_u=0.7,1.6$ (red solid and blue
long-dashed, respectively). 
Analogously to Fig. \ref{fig:t0gamma}, the inverted and 
isothermal models provide a better fit to the measured PDF. 
\emph{Central panel:} Starting from the isothermal
model ($\gamma_o=\gamma_u=1$, black dashed with errors), we vary the TDR index of 
overdensities to $\gamma_o=0.7$ (red solid) and $\gamma_o=1.6$ (blue
long-dashed).
The three curves are \review{very similar,} suggesting  that the 
regulated and transformed PDF is not significantly sensitive to 
the TDR of densities above the mean. \emph{Right Panel:} we show 
the PDF for three isothermal models with temperatures set to
$T_0=5000$ K (red solid), $T_0=15000$ K (black dashed with errors) and $T_0=25000$ K
(blue long-dashed). The differences are \review{small,} demonstrating the 
insensitivity of the PDF to the overall temperature of the IGM.}
\end{figure*}

We use a Markov-chain Monte Carlo (MCMC) technique to 
draw constraints from our data. 
For each \gdb{set of parameters}, we \gdb{define} a regular Cartesian
grid (specified in the next section) within the limits set by flat priors 
and evaluate the likelihood in eq. \ref{eq:likelihood}
at each point of the grid. The likelihood is then linearly interpolated
\gdb{between grid points}. MCMC chains are obtained
using the \gdb{{\sc EmCee}} package by \cite{MCpackage}.

\section{Results}\label{sec:results}

\gdb{We now present the results of fitting the PDF with the thermal models described in \S~\ref{sec:parametrizations} and summarized in Figure~\ref{fig:models_summary}.  }

\subsection{Models with a Standard $T_0$-$\gamma$ Parametrization}\label{sec:t0_gamma}

\gdb{As previous studies have demonstrated, in the  context of the}
 standard $T_0$-$\gamma$ parametrization 
the flux PDF is mostly sensitive to the slope $\gamma$. 
In Fig. \ref{fig:t0gamma} we show the  dependence of 
the PDF of the transformed flux on this parameter (solid lines), 
compared to the observed values (Black diamonds). We illustrate
the case of $\gamma=1.6,1.0,0.7$ in black, red dashed and blue long-dashed, respectively.
In this figure, the temperature $T_0=15000$ K is fixed, as well as the 
rescaling of the heating rates in the simulation (i.e. the 'smoothing 
parameter') $\xi=0.8$. The isothermal
($\gamma=1$) and the inverted ($\gamma=0.7$) models \gdb{are most similar to the} observed PDF, while the one with $\gamma=1.6$ 
produce a PDF that is significantly flatter.

\gdb{We run our}
MCMC analysis assuming the likelihood reported in eq.~\ref{eq:likelihood}
and the prior on the mean flux (eq.~\ref{eq:prior}). 
We use a regular Cartesian parameter grid where the parameters can assume
the following values: $\bar{F}\in \{0.7171,0.7271,0.7371,0.7471,0.7571 \}$;
$\gamma\in \{0.4,0.7,0.85,1,1.15,1.3,1.6,1.9\}$;
$T_0 \in \{5000,15000,25000,35000 \}$ ;
$\xi \in \{0.3,0.8\}$. \review{We choose small values for the smoothing parameters compared to the model that best match
the IGM temperature measured in B11. The choice
of a small $\xi$ is conservative as assuming higher smoothing would 
require lower} values of $\gamma$ in order to match the observed PDF, as  will
become clear when we later investigate  the degeneracy between $\xi$ and $\gamma$

 The results are summarized in Fig. 
\ref{fig:t0gammaconstraints}. We present the 68\% (dark red) 
and 95\% (light red) confidence levels in the planes defined by
the parameters $\gamma$-$T_0$ (left panel) and $\gamma$-$\xi$ 
(middle panel). In the right panel we show the full posterior 
distribution of $\gamma$, marginalized over all the other parameters. 
The horizontal dashed line marks the isothermal models ($\gamma=1$). 
All the models falling below this lines are called 'inverted'. 
\gdb{As see in Figure~\ref{fig:t0gammaconstraints}, an inverted model is preferred,} 
although an isothermal model is consistent at 
1-$\sigma$. This results is consistent with previous works 
that used the flux PDF  
\citep[e.g.][]{Bolton08,VielBolton09,Calura2012,Garzilli2012}, 
\gdb{and like} these it is in tension
with \gdb{constraints on $\gamma$ from} different techniques. In the following sections we explore possible 
models that could \gdb{resolve this} tension.

\subsection{Broken-Power Law Models of the Thermal State of the IGM}\label{sec:broken_t_rho}


We next examine broken power-law models of the thermal state of the IGM
described in \S~\ref{sec:parametrizations}.
In Fig. \ref{fig:broken_trho_pdf} we show the effect of the most
relevant parameters on the PDF of the transformed flux $F_{10}$. 
In the models shown in this figure we fix the break density to $\Delta_b=1$ (i.e. the mean
density), and the parameter regulating the smoothing to $\xi=0.8$. 
The leftmost panel shows the dependency on the index in \review{underdense regions} ($\gamma_u$).  
\gdb{A value of}
$\gamma_u=0.7$ \gdb{(inverted)} or $\gamma_u=1.0$ \gdb{(flat)} provides a much better 
fit than $\gamma_u=1.6$. 
The results  closely resemble those in  Fig. \ref{fig:t0gamma}, suggesting that the 
PDF is mostly sensitive to the temperature-density relationships at densities
below the mean. This is corroborated by the central \gdb{and right-hand} panels of figure
\ref{fig:broken_trho_pdf}, which explicitly demonstrate the small role
played by the \gdb{index in the overdense regions ($\gamma_o$) and the overall normalization of the temperature-density relation ($T_0$)} in modifying  the transformed PDF.

\begin{figure*}
\includegraphics[trim={1.5cm 0 2cm 0},clip,width=\textwidth]{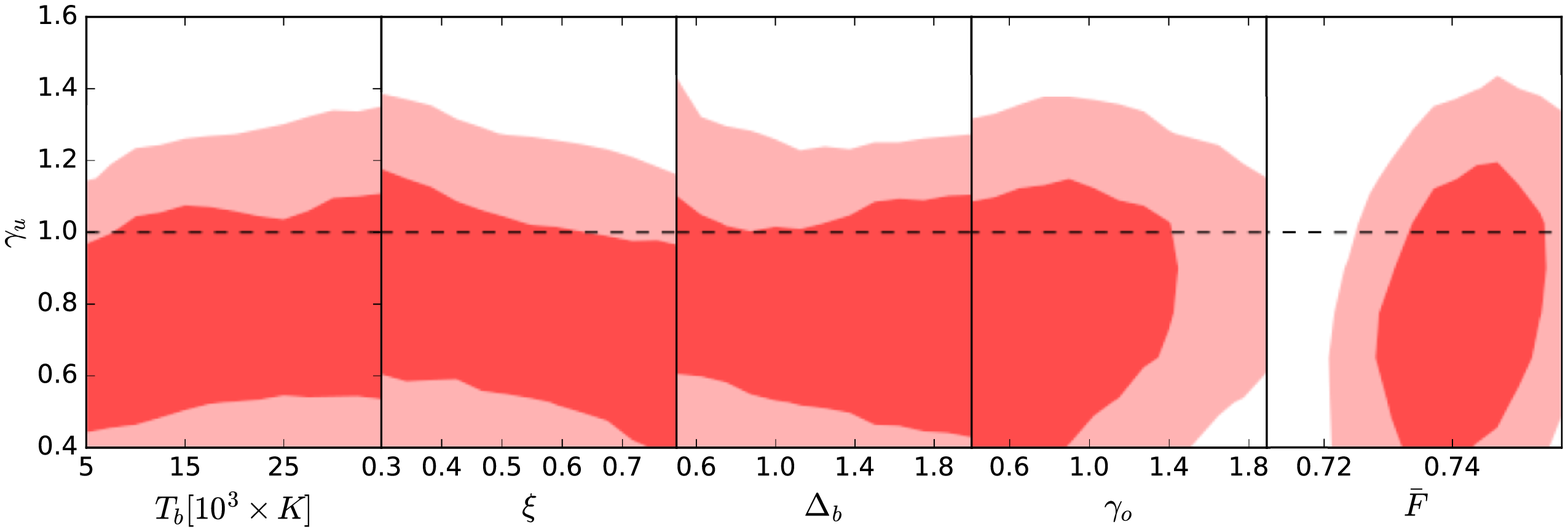}

\includegraphics[trim={0 0 0.1cm 0},clip,width=\textwidth]{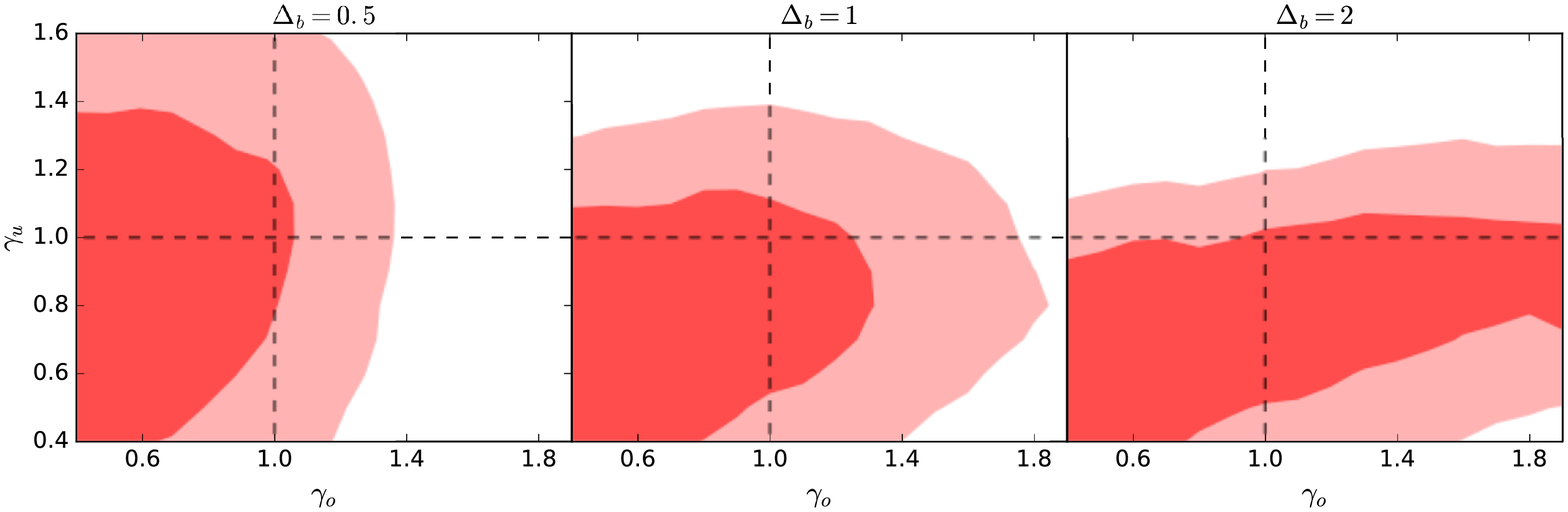}

\caption{\label{fig:gamma_u_contours} 
\review{\emph{Top row: }}Results of the MCMC run  applying the likelihood in 
eqn.\ref{eq:likelihood} to the PDF of the transformed flux
of the deep spectrum ($A=10$), in the parameter space of the
broken-power-law model. The plot shows the 68\%  (dark red) 
and 95\% (light red) confidence levels in the parameter subspaces 
defined by the power-law index of the TDR in \review{underdense regions} $\gamma_u$ and, from left to right, 
the break temperature $T_b$,
the smoothing parameter $\xi$, the break density
$\Delta_b$, the slope of the $T$-$\rho$ relationship in overdensities
$\gamma_o$  and the mean flux $\bar{F}$. A prior on $\bar{F}$ is 
adopted, as described in the text. 
The figure shows that, regardless of the choice of the other parameters,
an inverted or isothermal $T$-$\rho$ relationship is preferred in 
underdense regions. There is a mild degeneracy with $\xi$, which means
that if the pressure smoothing of the IGM is lower, slightly higher 
values for $\gamma_u$ are tolerated. 
\label{fig:fixed_delta_contours} 
\review{\emph{Bottom row:} the 1-$\sigma$ and 2-$\sigma$ 
confidence levels in the
$\gamma_u$-$\gamma_o$ plane as a function of the break density
$\Delta_b$. All other parameters are marginalized over.
The dashed horizontal (vertical) lines trace the points
where $\gamma_u=1$ ($\gamma_o=1$). 
The contours broaden in the $\gamma_o$
dimension as $\Delta_b$ increases, while $\gamma_u$ shows the 
opposite behaviour.  If  the 
break point is set to $\Delta_b=2$ (right most panel), no constraint
can be placed on $\gamma_o$, indicating that the transformed flux pdf
is not affected by the thermal state of densities above that value.} 
}
\end{figure*}

We carried out a full MCMC analysis in the parameter space defined by 
$\gamma_u,\gamma_o,T_b,\Delta_b ,\bar{F}$ and $\xi$, adopting the same
 prior on $\bar{F}$ as in the previous section. The range and spacing  
of the grid in  $\gamma_u$, $\gamma_o$ and $T_b$ is the same as those of $\gamma$
and $T_0$ in the previous section, and those of $\bar{F}$ and $\xi$ are 
unchanged. The break density  is \review{varied in the range} $\Delta_b\in\{0.5,1,1.5,2\}$.

The constraints 
are presented in Fig. \ref{fig:gamma_u_contours}. The most important
result is that regardless of the choice for the other parameters, 
the thermal index of \review{underdense regions} $\gamma_u$ is preferentially inverted or
isothermal, with $\gamma_u \gtrsim 1.2$ excluded at 2-$\sigma$.
The degeneracy of $\gamma_u$ with the other parameters is not strong,
although lower values of the pressure smoothing do allow a higher 
$\gamma_u$. 
There is also a slight preference for lower values of $\gamma_o$. 
\gdb{Note}, however, that since the 
break density $\Delta_b$ is a free parameter the physical meaning
of $\gamma_o$ and $\gamma_u$ varies as well.

\begin{figure*}
	\includegraphics[trim={0.7cm 0 0.1cm 0},clip,width=\textwidth]{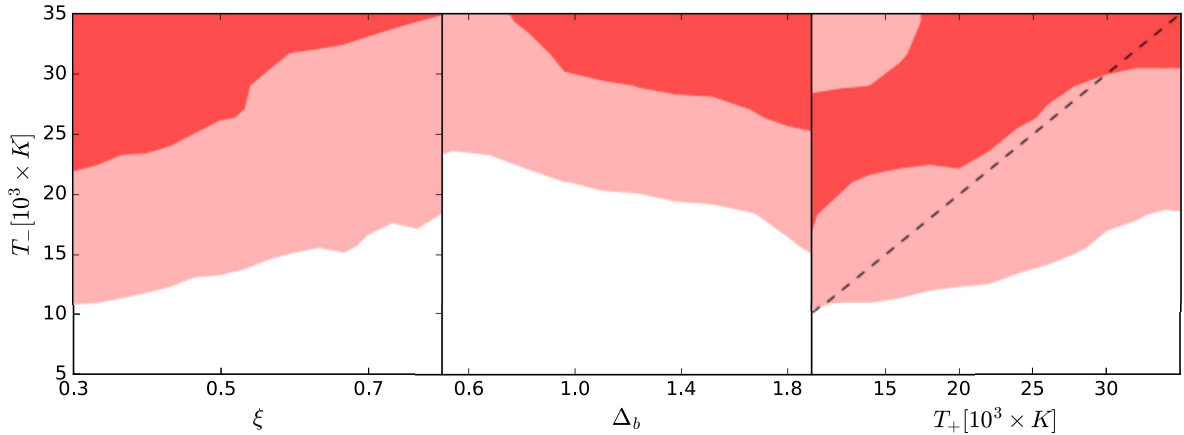}
	\caption{\label{fig:contour_step} This figure is analogous to Fig. 	
	\ref{fig:gamma_u_contours}, but in the parameter space of 
	the step-function model for the temperature-density relationship. 
	The figure reports the 1-$\sigma$ and 
	2-$\sigma$ contours in the plane defined by $T_-$ and, from left
	to right, the smoothing parameter $\xi$, the discontinuity density
	$\Delta_b$ and the temperature of overdensities  $T_+$. 
	The black dashed line in the rightmost panel traces the identity
	$T_-=T_+$. The figure, and in particular the right panel,  
	suggest that \review{underdense regions are relatively} hot. The required temperature in under-dense regions
	is comparable or higher then the temperature in overdensities, in the
	framework of this  simple two-temperature model}
\end{figure*}

\gdb{To illustrate the role played by the break density,} we ran three more MCMCs where we fix $\Delta_b$ to 0.5,1 and 2. For each run we then 
calculate the confidence levels in the $\gamma_u$-$\gamma_o$ plane.
The correspondent contours are shown in figure \ref{fig:fixed_delta_contours}.
As the break moves from low to higher densities (left to right)
there is a clear trend for the contours to shrink in $\gamma_u$ 
and expand in $\gamma_o$. We interpret this 
as the consequence of the different density ranges described by
$\gamma_u$ and $\gamma_o$ in the three cases. For example, in the 
right panel $\gamma_o$ only affects densities above $\Delta_b=2$. The 
fact that the distribution is so broad in $\gamma_o$ suggests that 
the transformed PDF is not sensitive to such overdensities. In the 
left plot, instead, the constraints on the two indices are relatively
similar, suggesting that $\Delta_b=0.5$ sits in the 
density range that prefers
an isothermal or inverted $T$-$\rho$ relationship.
The central plot
lies between the other two cases. 
\subsection{A Step-Function Model of  the Thermal State of the IGM}

\gdb{For the step function given by equation~(\ref{eq:step_function}),}
we generate a grid of models where we vary $T_+$ and
$T_-$ between 5000 K and 35000 K, $\Delta_b$ between 0.5 and 2, and
$\bar{F}$   and $\xi$ between 0.3 and 0.8 with the usual priors
and grid spacings. 

The most relevant results \gdb{are shown} in figure \ref{fig:contour_step}.
The contours show the constraints on  $T_-$ and its
degeneracies with the parameters $\xi,\Delta_b$ and $T_+$. 
The first \gdb{point to notice} is that cold underdense regions ($T_-<15000$ K) are excluded at 
2-$\sigma$, independent of the \gdb{values} of other parameters.
The degeneracy with the smoothing parameters $\xi$ suggests that 
\gdb{the preference for hot underdense regions could be partially mitigated by increasing}
the fluctuations on small scales \gdb{through a reduction of the} thermal pressure 
(analogously to \S~\ref{sec:t0_gamma}).

The left panel of Figure~\ref{fig:contour_step} illustrates the constraints on the contrast 
between $T_-$ and $T_+$. As a reference, we plot the identity
line $T_-=T_+$ as a dashed black line. \gdb{Nearly the entire} 
1-$\sigma$ region \gdb{falls above this} line, implying 
that the PDF demands $T_+>T_-$. This test 
\gdb{supports the general conclusions that} the temperatures in \review{underdense regions} need to be similar or higher
than those in overdensities.

\subsection{Temperature Fluctuations}

\begin{figure*}
\includegraphics[trim={0.7cm 0 0 0},clip,width=0.35\textwidth]{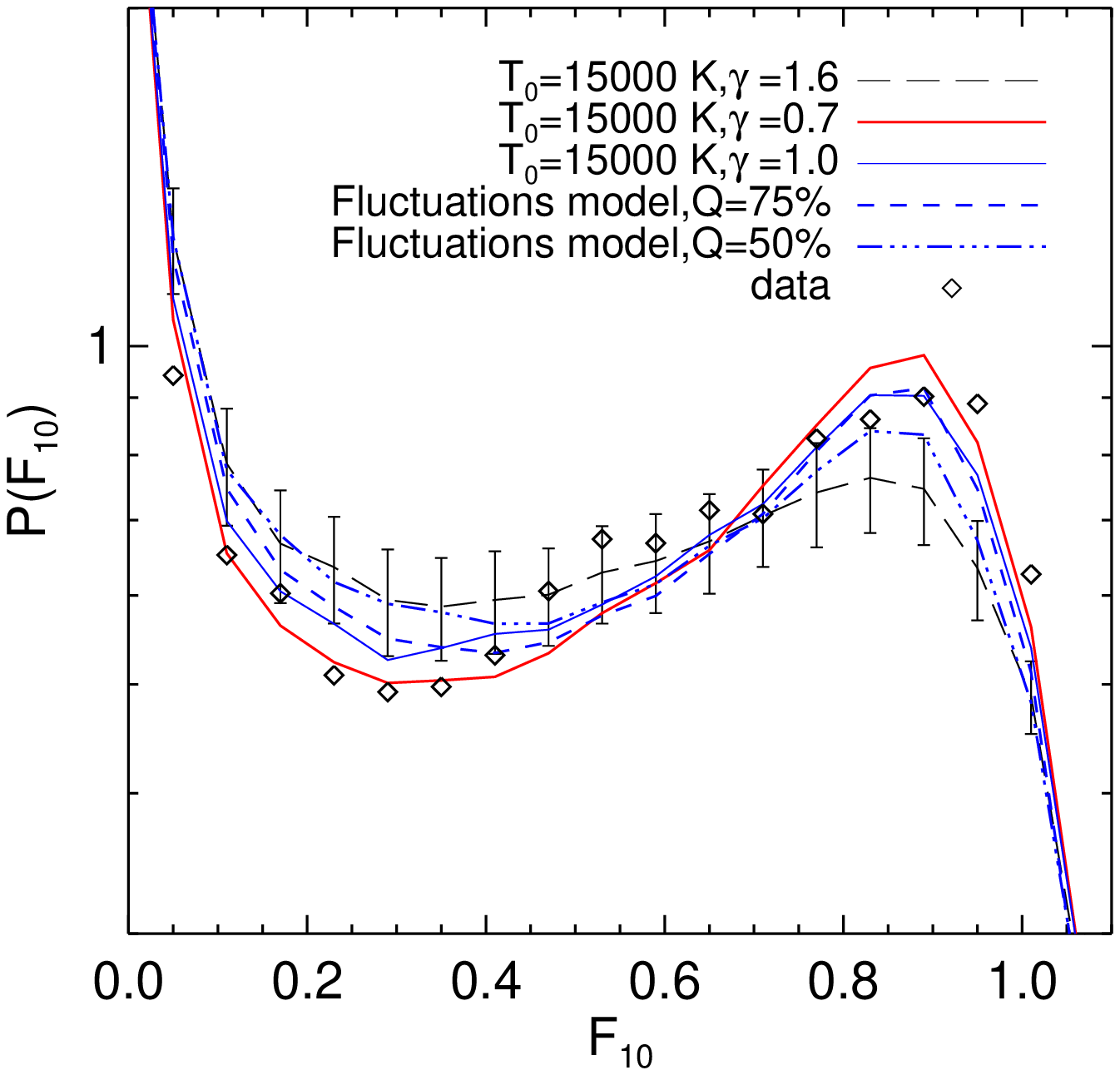}
\includegraphics[trim={0 0.1cm 0.5cm 0},clip,width=0.6\textwidth]{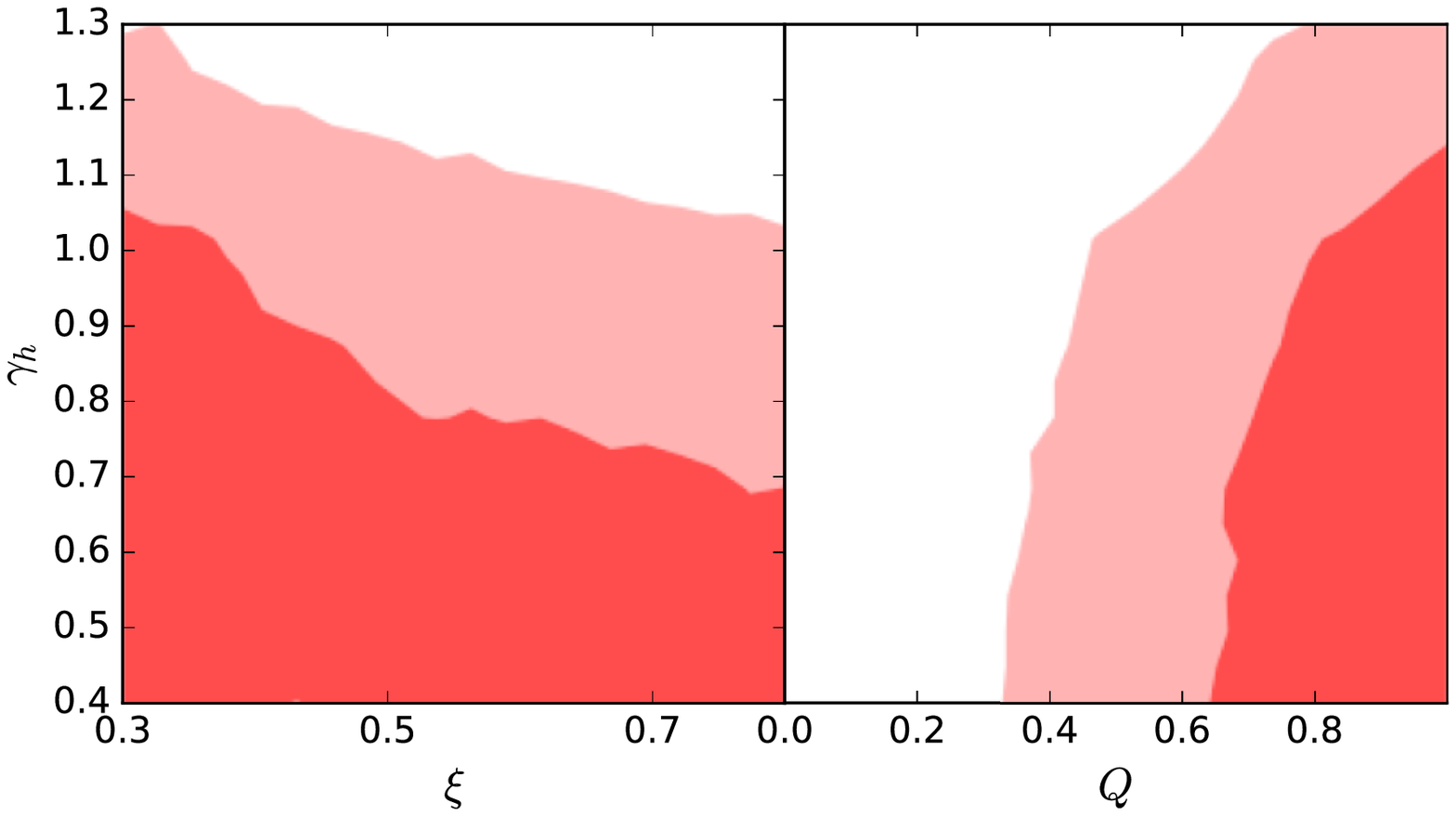}
\caption{\label{fig:fluctuations_PDF} 
\review{\emph{Left panel:} PDF} of $F_{10}$ from the DEEP
spectrum (black diamonds) compared to the prediction of  
a standard $\gamma=1.6$
model (black long-dashed), an inverted $\gamma=0.7$ model of the thermal 
state of the IGM (red solid) 
and two different models with temperature fluctuations. The 
temperature fluctuation models are built by splitting the simulation box in two parts, 
one of which has $T_0=5000$ K and $\gamma=1.6$ (cold component),
the other $T_0=35000$ K and $\gamma=1.0$ (hot component). The 
proportion of the two parts sets the filling factor of the 
hot component $Q=V_{\rm hot}/(V_{\rm cold}+V_{\rm hot})$. 
The figure shows
in blue the cases where the hot component occupies 75\% (dashed)
or 50\% (dotted-dashed) of the total volume. For comparison, we also show
the case of an isothermal IGM, with $T_0=15000$ K and $\gamma=1$ (blue solid). 
We find  that  models with a strong temperature contrast
and high filling factor ($Q>50$\%) provide a good fit to the 
PDF without need of an inverted TDR anywhere.
\label{fig:fluct_contours}\review{\emph{Right panel:} 68\% and 95\% confidence levels 
in the parameter space that defines the temperature fluctuation models.} 
$\gamma_h$ is the slope of the temperature-density relationships in hot 
regions, $\xi$ is the smoothing parameter and $Q$ is the filling factor of the hot bubbles.
The results have been  marginalized over the temperature of the hot bubbles $T_h$,
the temperature in the cold regions $T_c$ and the mean flux $\bar{F}$. }
\end{figure*}

\gdb{We finally turn towards our  simple  models for temperature fluctuations.  The impacts of the relevant parameters on the}
 transformed PDF are shown in Fig.
\ref{fig:fluctuations_PDF}. The figure shows the PDF of 
two 'uniform' (i.e. without fluctuations) models of the IGM with
$\gamma=1.6,1 $ and 0.7 (in black \review{long-dashed}, blue and red, respectively); the latter provides a 
better fit to the data (black diamonds) as shown in previous
sections. The blue lines represent two models of temperature fluctuations
with \gdb{the filling factor of the hot gas, $Q$, set to} 75\% and 50\% (dashed and dotted-dashed, respectively).  \gdb{Here} the temperature at mean density, $T_0$, is 5000 K in the cold regions and 
35000 K in the hot regions.  The hot regions are assumed to be isothermal 
and the mean flux \emph{of the entire volume} is always imposed to be
$\bar{F}=0.7371$ by rescaling the UV background.
\gdb{Note} that if the  filling factor is greater than $50\%$
then the transformed flux PDF becomes similar to the one obtained from an inverted model. 

\gdb{We emphasize} that the PDF of a model with temperature fluctuations is not simply a weighted average 
of the two models characterizing the hot and the cold parts. 
This would be the 
case only if the optical depth was not renormalized to match the mean
flux constrained by observations. In a mixed model, hot regions are more 
transmissive than cold regions, due to the lower recombination rate of
hydrogen. Compared to the case where the IGM is completely filled with
hot, isothermal gas, in a fluctuating medium the transmission of the hot
bubbles must compensate the opacity of the cold regions, therefore the
UV background must be adjusted such that they are more transparent. The 
opposite argument could be made for the cold component, which needs to be 
more opaque when mixed with hot gas. The adjustment of the UV background
\gdb{becomes stronger as the temperature contrast between the hot and cold regions grows.}
Since this mechanism will increase the number of \gdb{$F \sim 1$ pixels in hot regions and $F \sim 0$ pixels in cold regions}, the net effect on the flux 
PDF is non trivial.
For illustrative purposes we have chosen \review{rather drastic} temperature fluctuations, 
\gdb{though more realistic models (related to He II reionization, for example), could also be tested with the flux PDF in similar quality data.}

We have performed a quantitative parameter analysis \gdb{for the temperature fluctuations model}. 
We consider the space defined by the quantities $\xi, \bar{F},T_c,T_h,\gamma_h,Q$. 
We define the parameter grid imposing the cold regions to have $\gamma=1.6$,
as expected long after reionization events \citep{HuiGnedin97} and temperatures
in the set $T_c\in\{5000,15000\}$ K. Conversely, the hot regions have higher 
temperatures $T_h\in\{15000,25000,35000\} $ K and a flatter index 
$\gamma_h\in\{0.4,0.7,0.85,1.0,1.15,1.3\}$. The filling factor is set to 
the values $Q\in\{0,1/4,1/2,3/4,1\}$.  Note that the standard models 
with an inverted  temperature-density relationship are a subset of this
parameter space, after projection to $Q=1$. 
Figure \ref{fig:fluct_contours} presents the most relevant results. 
The two panels illustrate the degeneracies of $\gamma_h$, the thermal
index in hot bubbles, with the smoothing parameter $\xi$ and the filling
factor $Q$. It is clear that even in the context of temperature fluctuations,
hot regions must be filled with inverted or close-to-isothermal gas. 
The usual degeneracy with $\xi$ applies, confirming the results of the 
previous sections. Filling factors close to unity are preferred, although
only $Q<0.4$ is excluded at 2-$\sigma$. If the volume filling fraction is low,
however, lower  values of $\gamma$ are required in order to match
the transformed PDF. 

We stress that these results depend on the simplified description we assume
for temperature fluctuations, as well as on our priors on the temperature
of cold and hot regions (respectively, flat \gdb{prior between $T_{c}=5000$} and 15000 K,
flat \gdb{prior between $T_{h}=15000$} and 35000 K). More realistic and theoretically 
motivated models are needed in order to properly test this scenario. 
However, this simple exercise reveals that including fluctuations 
modifies the flux PDF in the direction required
by the \gdb{data.} 

\section{Comparison with Other Techniques}\label{sec:technique_comparison}

In this section we \gdb{examine} the consistency of our results with
other methods \gdb{that have been used} to constrain the
thermal state of the IGM using the \mlya\ forest. In particular we will show
that the different statistics are sensitive to different \review{density ranges.}
We first focus on the standard flux PDF and  
flux power spectrum.  \gdb{We then} review the line
fitting procedure, and finally demonstrate the consistency 
of our results with the recent work by \cite{KGpdf15} on the flux PDF
from the BOSS quasar sample.

\begin{figure*}
\includegraphics[width=\textwidth]{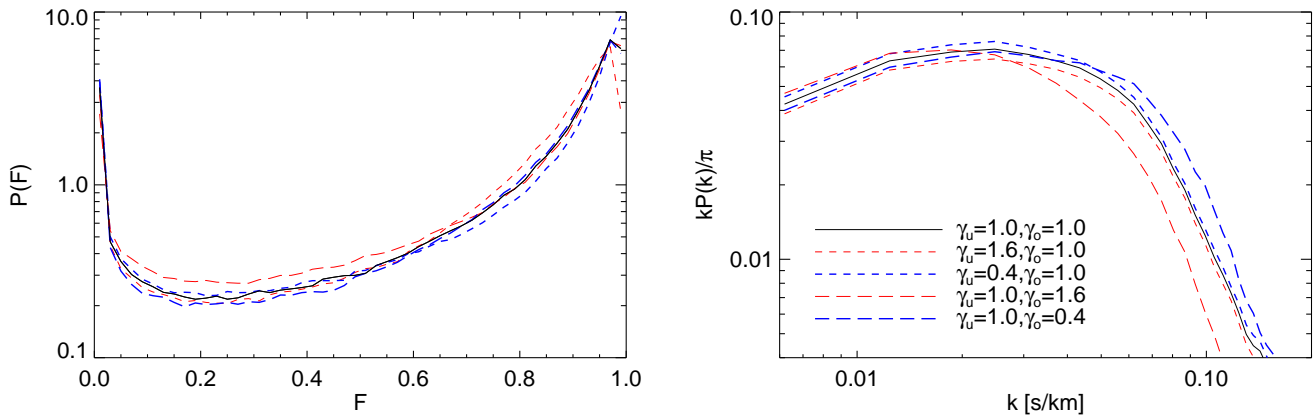}
\caption{\label{fig:pdf_vs_ps} \emph{Left panel:} the standard
flux PDF (\emph{not} regulated nor transformed) as a function of
the indices $\gamma_u$ and $\gamma_o$. The black solid lines represent
a model with \review{$\gamma_u=\gamma_o=1.0$ }. If we vary $\gamma_u$ to 1.6 
\review{(red dashed) or 0.4 (blue thick dashed)} the PDF changes around the peak,
at $F\gtrsim 0.6$, while the change occurs at low fluxes for 
$\gamma_o=1.6$ \review{(red long-dashed) and $\gamma_o=0.4$ (blue thick long-dashed)}.
\emph{Right panel:} the flux 1d power spectrum calculated from the
same set of models as the upper panel. The shape of the high-$k$ 
cut-off is only affected by the slope in 
overdensities (\review{long-dashed lines}),
while the effect of varying $\gamma_u$ (\review{dashed lines}) is
\review{a small overall shift} of the power at all scales. 
Techniques that rely on the position of the 
cut-off in the flux power spectrum to determine
the temperature are therefore measuring the IGM thermal state at 
$\Delta \gtrsim 1$. Conversely, the standard flux PDF is sensitive to 
both regimes, depending on the flux level. However, the shape of the 
peak ($0.6<F<1.0$) is mostly dependent on \review{the thermal state of
underdense regions}.  }
\end{figure*}
\subsection{The Flux PDF and Power Spectrum}

We have \gdb{shown} in \S~\ref{sec:broken_t_rho} that the PDF of the 
transformed flux $F_{10}$ is mostly sensitive to densities around and
below the mean. Here we 
\gdb{check whether} the same argument holds for 
the standard \gdb{(non-regulated, non-transformed)} flux PDF. Analogous to 
figure \ref{fig:broken_trho_pdf}, we consider a set of ``broken power-law'' models and 
we alternatively vary the TDR above and below the
mean density, observing how the PDF responds. \gdb{In all cases we fix $\Delta_b = 1$ and $T_b = 15000$ K.}  The results are shown 
in the top panel of figure \ref{fig:pdf_vs_ps}. The black line represents
a ``reference'' model where both indices are set to
$\gamma_u=\gamma_o=1.0$. \review{The long-dashed curves illustrate what happens when
the relation between temperature and density in overdensities is made
steeper (red, $\gamma_o=1.6$) or inverted (blue thick, 
$\gamma_u=0.4$)}. The PDF changes only at 
flux levels below $F\lesssim 0.6$, while the peak shape is left 
almost unchanged. The opposite happens when we vary $\gamma_u$ 
\review{(dashed lines)}, in which case most of the variation of the PDF occurs
around the peak. Therefore the PDF is in principle sensitive to a wide
range of densities. We note however that 
most of the forest pixels lie in underdense regions, i.e. in proximity of 
the peak, \review{hence we argue that this is the range to which the 
PDF is mainly sensitive.}
%

\gdb{The results are qualitatively different for the flux power spectrum, $P(k)$.  In the lower panel of Fig.~\ref{fig:pdf_vs_ps}
we show $P(k)$ computed for the same broken-power-law thermal models used above.  Here, we have computed the  power spectrum of the flux contrast, $\delta F$, defined as
\begin{equation}\label{eq:flux_constrast}
\delta F = \frac{F}{\langle F \rangle} - 1 \, ,
\end{equation}
where $\langle F \rangle$ \review{is the mean flux.
The shape of the cut off \gdb{in $P(k)$} changes only if $\gamma_o$ varies
(dotted-dashed lines), while the net effect of a variation in $\gamma_u$ 
is just a slight overall renormalization (dashed).}
\gdb{This makes sense as the} sensitivity of the 
cutoff shape is dominated by the thermal broadening of absorbers identifiable as 
lines, while the normalization is set by the amplitude of flux 
fluctuations at all scales. While the former depends on the temperature
at mild overdensities, as it will be obvious from the next section, the
latter follows the \review{full} mapping of density to \mlya\ flux. }

\gdb{By extension, we infer that any method} that uses the smoothness of absorption lines 
as a proxy for the temperature \gdb{at these redshifts} will most likely be sensitive to the thermal
state of high-density \gdb{regions.  Such techniques} include 
the power spectrum, wavelet
analysis, curvature and line-fitting.  In the next section we examine the \gdb{line-fitting} method explicitly.

\subsection{Voigt-Profile Fitting of Absorption Lines}

The line-fitting approach \gdb{decomposes the \mlya\ forest into individual lines fit by Voigt profiles.}
\gdb{The fits are then used to construct a}
2-d distribution of the H\,I column densities, $N_{\rm H\, I}$, and Doppler parameters,
$b$, of the absorbers. It has been \gdb{shown \citep{Schaye1999} that the position and slope of the lower envelope}
in the $\log N_{\rm H\, I}$-$\log b$ plane is
closely related 
to the thermal parameters $T_0,\gamma$. Measurements based on this
idea, \gdb{most recently by \citet{Rudie2012} and \citet{Bolton14},} have led to results that are clearly in contrast with a 
$\gamma<1$ TDR. 

\begin{figure*}
\includegraphics[width=\textwidth]{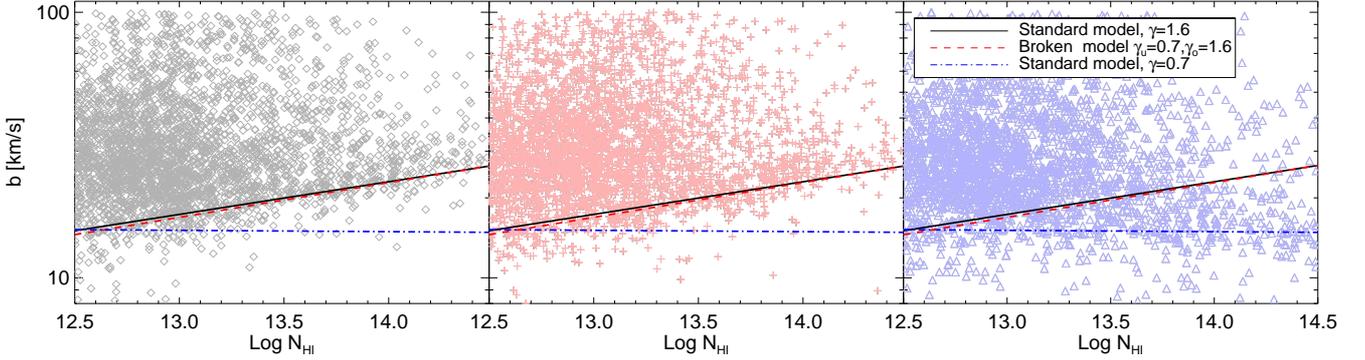}
\caption{\label{fig:nb_dist}
\review{The line distribution in the $b$-$N_{HI}$
plane. The symbols in light colors correspond to 
individual fitted lines in three IGM models: a}
standard IGM model with $\gamma=1.6$ and $T_0=10000$ K \review{(black,
left panel), a broken power-law model with $\gamma_u=0.7,\gamma_o=1.6
,\Delta_b=1,T_b=10000$ K (middle panel, red) and
a standard model with $\gamma=0.7$ and $T_0=10000$ K (blue,
right panel).} The lines 
represent the fitted power-law to the lower cut-off of the distribution.
The fit is done using the algorithm by \protect\cite{Schaye00} and the lines are
matched by color to the models they are derived from. 
The \review{left and the middle panels show that the line distribution is barely
changed if the thermal state of the IGM is modified only in 
underdense regions.} 
Conversely (right panel), the shape of the cut off is highly sensitive
to the temperature-density dependency in overdensities. We therefore
conclude that this technique probes the IGM at densities higher than the mean, differently than the PDF.}
\end{figure*}

Following  \citet{Bolton14} we argue that this method \gdb{preferentially} probes the temperature-density relationship at overdense gas. 
To \gdb{demonstrate this} we consider two models. One is a \gdb{``standard''}
model with $T_0=10000$ K and $\gamma=1.6$. The other one is 
a model with \review{inverted underdense regions,} i.e. with a break at $\Delta_b=1$ and
 $\gamma_u=0.7$. By setting $T_b=10000$ K and 
$\gamma_o=1.6$ we ensure that the two models are 
indistinguishable above the mean density. As a consistency
check, we also analyse a standard model with  $T_0=10000$ K
and $\gamma=0.7$. 

We then generate a mock sample of spectra for  the models,
which are forward-modelled as in \S~\ref{sec:method}. 
{\sc vpfit} is run \gdb{on} the three datasets  to obtain the
$N_{\rm H\, I}$-$b$ distributions plotted in Fig.~\ref{fig:nb_dist}.
Following \citet{Bolton14} we consider in the analysis only lines with
$12.5<\log N_{HI}<14.5$ (in cm$^{-2}$), $8<b<100$ (in km/s) and 
with relative error on $b$ lower than 50\%. 
We use the algorithm described  in \cite{Schaye00} to fit the lower
cutoff of the distributions (coloured lines in the plot), 
assuming a power-law relation 
\begin{equation}
b=b_0 (N_{HI}/10^{12} cm^{-2})^{\Gamma-1}.
\end{equation} 
\gdb{For the standard and broken models we find} 
$\Gamma_{\rm standard}-1=0.146^{+0.033}_{-0.009}$ and 
$\Gamma_{\rm broken}-1=0.131^{+0.052}_{-0.031}$. 
In \citet{Bolton14} they assume the relation 
$\gamma = 1+ \xi_2/\xi_1(\Gamma-1)$, where $\xi_1$ and 
$\xi_2$ are determined from an hydrodynamical simulation, giving
$\xi_2=2.23$ and $\xi_1=0.65$. If we follow this assumption, our fit 
translates into a ``measurement'' of gamma in these two
models of 
$\gamma_{\rm standard} = 1.50^{+0.11}_{-0.03}$ and 
$\gamma_{\rm broken} = 1.45^{+0.17}_{-0.10}$ 
All errors quoted
are 1-$\sigma$. The estimates of $\gamma-1$ in the two models are 
consistent with each other, and consistent with power-law index of 
overdensities $\gamma_o=1.6$.
This test demonstrates that line-fitting techniques \gdb{do indeed} probe the slope of the
temperature-density relationship in overdensities, \gdb{and are largely insensitive to the thermal properties of the \review{underdense regions}.}

\subsection{The BOSS Flux Probability Density Function}

Recently, \cite{KGpdf15} calculated the \mlya\ flux PDF from a 
large set of quasars from the BOSS sample and discussed in detail the
implications for the thermal parameters. They find that an isothermal
or an inverted temperature-density relationship is not statistically 
consistent with their measurement at any redshift. In this section we
present a possible explanation for the tension with our results. 
The data set employed in their analysis is  highly complementary to ours:
a large \review{ sample of spectra with moderate signal-to-noise ratio and lower resolution} compared to a single spectrum with extremely high signal-to-noise ratio.
\review{We will investigate now whether the PDF of moderate signal-to-noise and 
low-resolution data also probes the regime where
absorption features are more prominent, i.e. at densities higher than the mean.}

To assess this quantitatively, we 
analyse two sets of mock data, assuming that the IGM
is described by a broken model as used in \S~\ref{sec:broken_t_rho}.
The fiducial model we use has $\gamma_u=1$,$\gamma_o=1.6$,$\Delta_b=1$
and $T_b=15000$ K.
As a consistency check, we perform the same analysis on a standard 
model with $T_0=15000$ K and $\gamma=1.6$. 

\gdb{For each model,} we generate 
a mock sample with the spectral
properties of the BOSS data, as in \cite{KGpdf15}, and of the 
DEEP spectrum. 
In particular we assume that the BOSS-like sample has a path length 
of 500 \mlya\ forest regions, resolution of R=2000, pixel scale $\Delta v=69$
km/s and a  signal to noise of $\sim 10$ per pixel. 
The DEEP spectrum-like mock sample is constructed 
as described in \S~\ref{sec:forward_modelling}.  \gdb{We then apply our MCMC analysis to the four flux PDFs (in this case not transformed or regulated) generated from the mock data.}

For this test we consider only the bi-dimensional parameter space defined by 
$T_0$ and $\gamma$. Note that this space does not include the 'true' model 
in the case where \gdb{a broken power-law is used.}
This is done on purpose in order to understand what bias we may get 
with a wrong parametrization, depending on the properties of the
data. 

\begin{figure*}
\includegraphics[width=\textwidth]{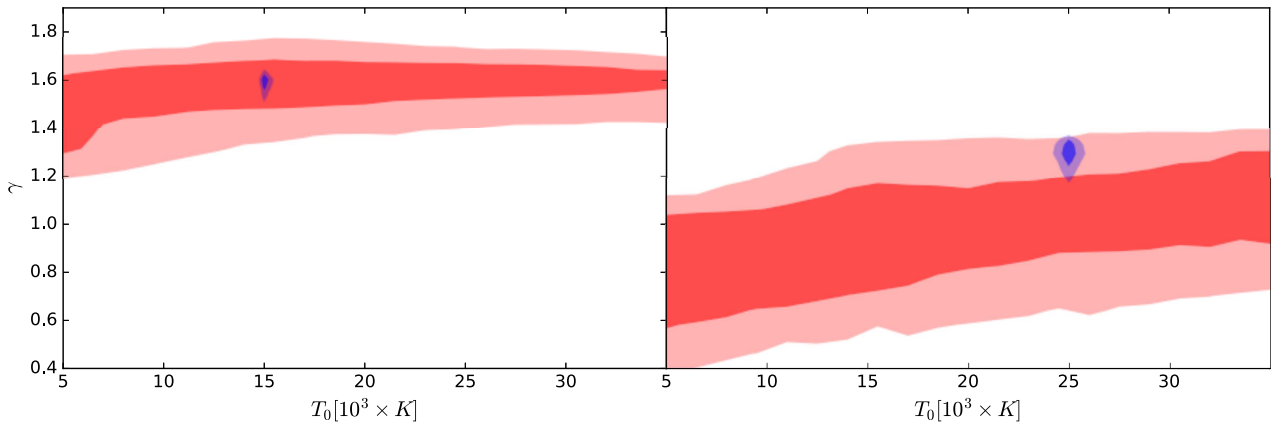}
\caption{\label{fig:pdf_standard} 
Parameter study of two mock samples of spectra generated from a 
standard model with $T_0=15000$ K and $\gamma=1.6$ (left panel)
and on a broken model with $\Delta_b=1,T_b=1500 K,\gamma_o=1.6$
and $\gamma_u=1$ (right panel). In both cases the smoothing 
parameter is $\xi=0.8$. One sample is built assuming the path length,
signal to noise, pixel scale and resolution of the DEEP spectrum, the other 
mimicking the path length, noise, pixel scale and resolution of 
the BOSS sample analysed by \protect\cite{KGpdf15}. The pdf of the two 
dataset is calculated for the two  models of the thermal state, and then analysed
using a set of models where \emph{only $T_0$ and $\gamma$ are 
varied}. The confidence levels are then obtained through the 
standard MCMC technique, shown in red (DEEP-like sample) 
and in blue (BOSS-like sample). The left panel simply confirms that
the correct value of $T_0$ and $\gamma$ are recovered, and reveals 
the lack of sensitivity of the PDF from high-quality data  to $T_0$. 
The high precision of the BOSS sample is due to the large path length
of the data. The right panel illustrates the bias induced on the thermal
parameters if the 'true' IGM is described by a broken model instead of 
a standard one. A sample with high signal to noise like the DEEP 
spectrum would be strongly biased towards the values of $\gamma$ in
the \review{underdense regions} ($\gamma_u=1$ in this example), while a noisy and large 
sample like BOSS would return an intermediate value between 
$\gamma_o$ and $\gamma_u$. We interpret this result as a hint that noisier
data are less sensitive (in a relative sense) to low densities than
high-signal data. }
\end{figure*}

The results of this test are shown in figure \ref{fig:pdf_standard}.
On the left we report the results of the consistency check, 
where we fit
$T_0 $ and $\gamma$ for the fiducial model with 
$T_0=15000 $ K and 
$\gamma=1.6$. Thanks to the large sample size, the BOSS-type data
are able to recover the correct parameter values with high precision.
The result for the DEEP sample, constructed with the path length of one spectrum, is 
less accurate but is consistent with the fiducial value. It is also
interesting to note the absence of a degeneracy with $T_0$.

\gdb{The case of a broken power law} is illustrated on the right.
The precision of both samples is significantly degraded. The constraints
derived from the DEEP mock sample sit around $\gamma=1 $, i.e. the \gdb{value of $\gamma$ in} \review{underdense regions,} and a slight degeneracy  with $T_0$ appears. 
The BOSS mock instead ``measures'' a value of $\gamma \approx 1.3$, which is 
intermediate  between the two indices $\gamma_u=1$ and $\gamma_o=1.6$. 
\gdb{From this we infer that the high resolution, high signal-to-noise data are more sensitive to lower densities, as argued above, while the \review{PDF of the moderate-signal-to-noise}  BOSS data are also sensitive to higher densities.  This may partly explain why \cite{KGpdf15} favored values of $\gamma \sim 1.5$, while previous PDF analyses using high-resolution data tended to favor lower values of $\gamma$.}
We stress that in this particular test we are assuming a 2d parameter space ($T_0$-$\gamma$), for sake of simplicity. This means that the 
constraints from the the mock data are more precise
than what is  achievable with real data in a full analysis. 

\section{Discussion}\label{sec:discussion}
We \gdb{have attempted here to shed light on the apparent}
tension between constraints on the thermal state of the IGM achieved with
different techniques. The controversial
results yielded by the flux PDF have \gdb{previously} been debated in the context of systematic
errors. In particular, some papers showed that misplacing the continuum
level could lead to a constraint on $\gamma$ towards lower values \citep{Lee2012}, or that 
errors on the flux PDF were underestimated \citep{Rollinde2013}, \gdb{exaggerating}
the statistical significance of the claimed inverted temperature-density
relationship. Here we paid particular attention to the problem of continuum 
fitting, by splitting the data into 10 Mpc$/h$ chunks and by applying
\gdb{flux regulation based on the (relatively stable)} 95th flux percentile, as described in \S~\ref{sec:continuum_regulation}.
A test described in \review{appendix}~\ref{sec:continuum_test_mocks} shows how robust the PDF
is to continuum uncertainty after this regulation is applied. The only reason
the continuum could nevertheless be a source of bias would be if the shape
of emission lines in QSO spectra is really different than our assumptions, 
in particular if the characteristic scale of continuum variations was significantly smaller than 
1000 km/s. Given what we know about  the 
characteristics of quasar spectra redwards of the \mlya\ line,
and at lower redshift, where the forest is highly transmissive 
\citep{Scott2004}, this appears unlikely.
Furthermore, such small scales fluctuations of the continuum would be 
problematic for any \mlya\ forest statistic, not just the PDF. 

Concerning the estimation of our covariance matrix, with only one spectrum we
\gdb{are unable to estimate the covariance}
via bootstrap or jackknife techniques.
\cite{Rollinde2013} have shown that such methods may underestimate the variance unless
the chunks on which the dataset is resampled are larger than $\sim 25$ \AA .
As explained in \S~\ref{sec:error_estimation}, \gdb{however,} we used \gdb{simulated data} to
calculate the covariance matrix for a forest path length equivalent to the 
DEEP spectrum. Hence, the question of the accuracy of our errorbars is 
really a question about the convergence of our simulation with regard to box size. We have performed 
a convergence test which is described in detail in appendix \ref{app:convergence}.
\gdb{Using} two larger boxes of 20 and 40 Mpc$/h$, we find that 
the PDF is converged well within the estimated errors, while \review{variances  
are at most about 44\% larger in the 40 Mpc/$h$ box compared to the default run.
In our analysis we have therefore \gdb{increased} all the elements of the 
covariance matrix by 44\%}. We carried out a further
convergence test by using larger boxes from a recent set of simulations (Sherwood simulation\footnote{http://www.nottingham.ac.uk/astronomy/sherwood/},
Bolton et al, submitted), extending the box size to 80 and 160 Mpc$/h$. The errors
on the PDF are converged to $\lesssim 5\%$ for the 160 Mpc$/h$ box.  

We also argue that our approach of normalizing the continuum to the 95th percentile of the 
distribution makes the PDF less subject to large-scale variations. A simple
\gdb{explanation}, exact in the linear limit, is to consider large-scale fluctuations as stochastic 
changes in the mean density \review{along} the ``local'' 10-Mpc$/h$ 
\review{sight line}, i.e. all densities
are shifted by some constant. 
The effect on the forest would be a global 
renormalization of the transmitted flux, to which the PDF we use is  
invariant thanks to the regularization. We have in fact verified that not using 
the percentile regulation significantly degrades the level of convergence
with regard to the box size.

\section{Conclusions}\label{sec:conclusions}
We have investigated the thermal state of the low-density IGM using the 
\mlya\ forest of a single quasar spectrum with an exceptional signal-to-noise ratio of 280 per resolution element.  As part of the analysis, 
we have introduced a new technique to address  uncertainties 
due to continuum fitting. We   renormalise the flux distribution 
to a value close to the peak of the flux PDF  ( 95th percentile 
of the flux distribution). We apply
this 'flux regulation' to  chunks of \mlya\ forest 
of 10 (comoving)  Mpc$/h$, in order to  account for continuum variation on larger scales. 
Our analysis further employs  a  rescaling of the \mlya\ optical
depth. This gives more emphasis to the high transmission  
part of the \mlya\ flux PDF and allows us to better probe the low-density  IGM. 
Some information at the high transmissivity end is naturally lost due to the continuum regularization; however, the resulting PDF is virtually free from bias due to continuum placement errors (Fig. \ref{fig:pdf_rescaling}).   
Critically, the shape of the PDF of the regularised flux is still 
very sensitive to the TDR.
%
We have compared the flux PDF obtained from the  DEEP spectrum with the predictions
of a number of thermal models of the IGM, obtained by imposing different 
temperature-density relationships on two hydrodynamical simulations with fixed density and velocity fields. We have considered four different 
parametrizations for the TDR, starting with  the standard simple 
power law relation characterised by $T_0$ and$\gamma$. We then have moved to  more complex parametrisation 
with broken power laws and steps in the TDR,  to see whether in this way  the tension between the 
constraints on the thermal states estimated from the PDF and from 
other statistics of the \mlya\ forest could be reconciled.

For each of these parametrizations we have used Bayesian MCMC techniques to
obtain constraints. 
Our main results can be summarized as follows.

\begin{itemize}
	\item In our analysis of the  
	flux  PDF based on the  regularised and rescaled flux we find, assuming a power
	law TDR,  that $\gamma>1$ is excluded at \review{90\% confidence levels} and
	$\gamma>1.1$ at \review{95\% confidence level, after marginalization over
	the remaining parameters.} This results is consistent with previous
	measurement obtained from the PDF, but in contrast 
	with other methods. The fact that we still find this for our very high SN spectrum 
	and with our continuum regulation suggests that this discrepancy  is not solely 
	attributable to errors in the continuum placement. 

	\item We have tested a 'broken'  power law TDR in which densities 
	below and above a threshold $\Delta_b$ have different power-law indices 
	$\gamma_o$ and $\gamma_u$. Our constraints for this model suggest  
	a flat or inverted TDR around and below the mean density 
	($\Delta \lesssim 2$) provide a good fit to the measured PDF. 
	
	\item If the thermal state is assumed to be characterized by 
	two temperatures $T_+$ and $T_-$, assigned to densities above and 
	below $\Delta_b$, then models with $T_+ \lesssim T_-$ are favoured.
	This suggests  that the evidence for relative hot gas in \review{underdense regions} is robust with
	respect to the particular choice of the parametrization.  
	
	\item The fit to the observed PDF is also improved by our model 
	of temperature fluctuations as could be  possibly induced by HeII reionization.
	Our MCMC analysis for this model suggests again  that a significant fraction 
	of the volume ($Q>0.4$) must be filled with hot gas characterized
	by a flat or inverted TDR.	

	\item Regardless of the parametrization chosen, low values for the pressure 	
	smoothing parameters
	$\xi$ provide a better fit to the PDF and broaden the allowed range
	for the TDR parameters. This result is consistent with independent
	constraints on the smoothing from quasar pairs (Rorai et al, submitted).

	\item We \review{have shown  that the flux power spectrum and the 
	lower Doppler parameter cut-off  in the $N_{HI}$-$b$ distribution are 
	mainly sensitive  to the thermal state of the IGM at mean density and above,
	differently than the high-transmissivity part of the PDF that we analyse 
	here. Hence, the flux  PDF on one side and the 
	 the power spectrum and the line-fitting	
	techniques on the  other side are primarily 
	sensitive to sufficiently different densities that their measurements 
	are not actually in tension if a more flexible parametrisation is chosen 
	to characterise the thermal state of the IGM. }	
\end{itemize}	
Concerning the degeneracy between $\gamma$ and the smoothing parameter $\xi$,
it is interesting to note that a  recent analysis of Rorai et al. (2016, submitted) of the flux correlation 
in QSO pairs provided a first direct estimate of pressure smoothing. 
As Rorai et al. point out  the density field may also be different than that in  our simulations 
due to the effect of primordial magnetic fields or a different small scale matter 
power spectrum than we have assumed here.
Although the amplitude of the matter power spectrum on intermediate scales as characterised by $\sigma_8$ is now known with 
reasonably high precision \citep{Planck2015}, 
no constraints are yet available for  the matter  power spectrum on scales below $\sim 1$ Mpc. 
If the matter power spectrum is different than we assumed on 
small scales, this would certainly affect the \mlya\ forest statistics 
and in particular the PDF which is more sensitive to low-densities
in the linear and quasi-linear regime.	
%

The discrepancy of the flux PDF from that  expected if photo-heating of hydrogen dominates has been previously  
attributed to additional  heating sources like blazars  or to non-equilibrium and 
radiative transfer effects during HeII reionization. 
Our  analysis of the flux PDF of the \mlya\ forest region in the
ultra-high SN spectrum of HE0940-1050 confirms that, regardless of the particular 
parametrization we choose,  for a significant ($\gtrsim $ 50\%)  volume fraction of the  IGM the temperature 
in under dense regions appears  to be at least
as high  as the temperature at mean density and above. Both incomplete helium reionization
and blazar heating appear to be broadly consistent with this result and more detailed simulations 
will be needed to investigate if this still holds if either process is simulated more realistically.

Finally, we  emphasize that the results presented in this paper were obtained from a single
sight line. The next obvious step would be to extend our 
methodology to a larger sample of quasars. Although the signal to noise
level of the DEEP spectrum is unmatched, our results suggest that more standard high-quality high S/N data  
may nevertheless yield precise constraints for a wide range of 
densities. We further advocate in particular
the use of multiple and complementary \mlya\ forest
statistics, in order to discriminate
between the different thermal models of the IGM that were 
presented in this paper.

\section*{Acknowledgements} 
We thank Volker Springel for making GADGET-3 available.  This work made use of the DiRAC High Performance Computing System (HPCS) and the COSMOS shared memory service at the University of Cambridge. These are operated on behalf of the STFC DiRAC HPC facility.  This equipment is funded by BIS National E-infrastructure capital grant ST/J005673/1 and STFC grants ST/H008586/1, ST/K00333X/1.  We acknowledge PRACE for awarding us access to the Curie supercomputer, based in France at the Tres Grand Centre de Calcul (TGCC), through the 8th regular call. Support by the ERC Advanced Grant 320596
''The Emergence of structure during the epoch of reionization`` is
gratefully acknowledged.
ET is supported by the Australian Research Council Centre of Excellence for All-sky Astrophysics (CAASTRO), through project number CE110001020.
AR thanks Joseph F. Hennawi and the ENIGMA group at the Max Planck institute for Astronomy for helpful comments and discussion.
TSK acknowledges funding support to the European Research Council Starting 
Grant ``Cosmology with the IGM" through grant GA-257670.
PB is supported by the INAF PRIN-2014 grant "Windy black holes combing galaxy evolution". 
\bibliographystyle{mnras}
\bibliography{deepbiblio}

\appendix

\section{Sensitivity to the Noise and Resolution Estimates}\label{appendix:noise}
\begin{figure*}
\begin{center}
\includegraphics[width=\textwidth]{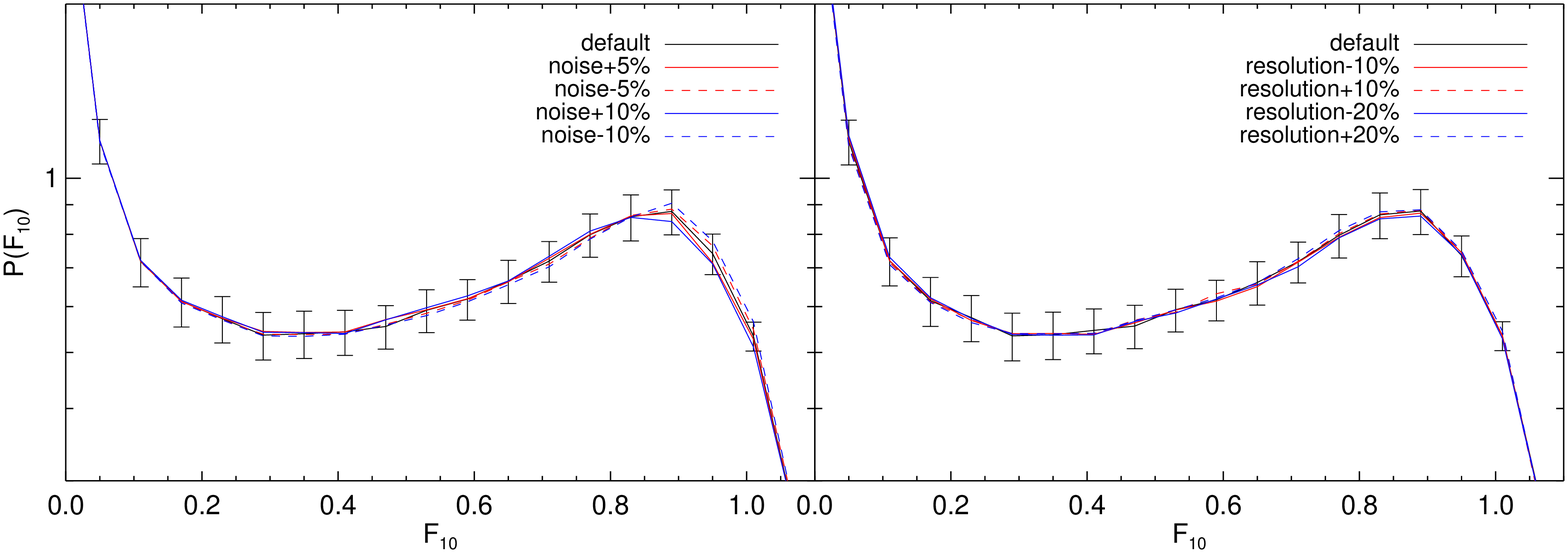}
   \caption{\label{fig:noise}
   \emph{The effect of varying the assumptions with regard to noise and resolution:} 
   in the upper panel we plot the transformed and regulated PDF calculated 
   from a model which uses the noise level estimated from the 
   pipeline (black solid) or in the case where it is increased 
   (solid lines) or decreased (dashed
   lines) by 5\% (red) and 10\% (blue); in the lower panel we show an
   analogous test where we compare the PDF computed assuming the 
   default resolution of 7.2 km/s (black solid) with the cases where
   the resolution is increased (dashed) or decreased (solid) by
   10\% (red) or 20\% (blue). All the lines in this plot are derived
   from a thermal model with $\xi=0.8$,$\gamma=1,T_0=15000$ K and
   $\bar{F}=0.7371$. The changes induced on the PDF are in all cases
   much smaller than the statistical errors calculated for the default
   model. 
   }     
\end{center}
\end{figure*}

The transformation of the optical depth defined in
\S~\ref{sec:transformation} amplifies the noise in proximity of the
continuum. We therefore need to demonstrate  that the transformed PDF
of the \mlya\ flux is not strongly sensitive to the noise and resolution
modelling when we employ an optical-depth rescaling factor as high as $A=10$. 
We verify this by calculating the PDF after varying our assumptions 
with regard to  noise level and resolution. We tested variations of 
5\%-10\% for the noise level and of 10\%-20\% for the spectral
resolution. The results are shown in figure \ref{fig:noise}.
The test is performed for a single model with $\xi=0.8$,$\gamma=1,T_0=15000$ K and $\bar{F}=0.7371$. The differences between the PDF of the  modified
and the default models are in all cases much smaller than the statistical
errors. We argue that this is a consequence of the extremely high signal to noise ratio of the spectrum, and of the fact that the 
\mlya\ absorption features are fully 
resolved at the spectral resolution of UVES.

\section{Mock Test of the Continuum Regulation}\label{sec:continuum_test_mocks}

\begin{figure*}
\includegraphics[width=\textwidth]{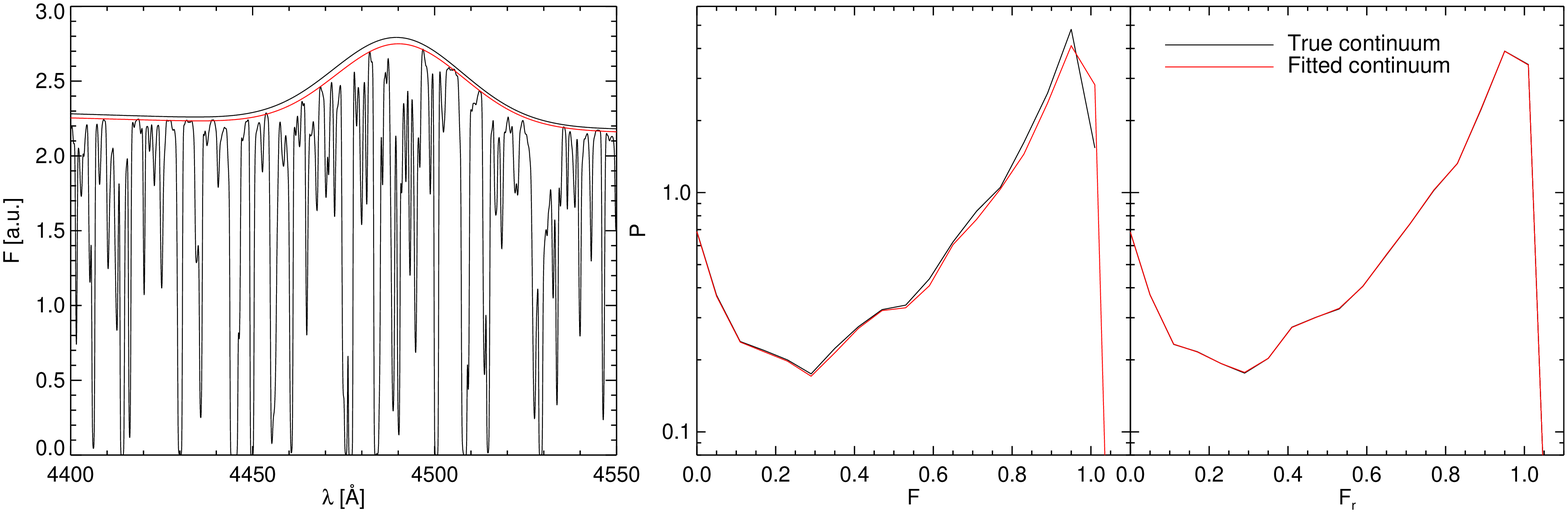}
\caption{\label{fig:spectra_cont_test} \review{\emph{Left panel:} example of a mock spectrum used
in our continuum regulation test (flux in arbitrary units). 
The true continuum of this spectrum is plotted as a black 
line, while the fitted version is shown in red. The fitted continuum
is systematically lower than the original one, but the slope and the
shape of mock emission lines are recovered with good accuracy. \label{fig:pdf_cont_test} \emph{Middle panel:} the 
standard flux PDF of the \mlya\ forest of the sample of mock  spectra 
assuming the true continuum (black) or the fitted one (red). The
continuum misplacement clearly results in an overestimation of the PDF 
toward high values of the flux,
as a consequence of using the high transmission regions as anchor points
for the fit. \emph{Right panel:} the same as the middle panel but 
using the percentile-regulated flux. The two PDFs are now practically identical, suggesting that 
on scales of 10 Mpc$/h$, to which our procedure is 
applied, the continuum error is well approximated by 
a normalization uncertainty.}
}
\end{figure*}

The 95-percentile regulation ensures invariance to a rigid 
rescaling of the continuum of the 10-Mpc$/h$ chunks. 
In reality, the true continuum 
could differ from the fitted one 
\gdb{in more complicated ways, particularly}
if emission lines are present. 
To assess the \gdb{robustness of our regulation} method, we perform the following test 
employing mock data.
\begin{itemize}
	\item We generate \gdb{10 mock \mlya\ forest spectra} with path length comparable
	to the real spectrum. For this we have concatenated  the (periodic) simulated spectra
	at the location of local minima and maxima, such such that the flux level at the juncture 
	differs by less then 0.01. This ensures that the spectra are 
	practically $C^1$ continuous.   
	
	\item Each of these spectra are multiplied by a continuum level with
 	a slope defined by $C\propto z^{\alpha}$, where $z$ is the \mlya\ 
 	absorption redshift and $\alpha$ is randomly chosen
 	between -1.2 and 0.3 . We randomly add between zero and five \gdb{Gaussian} emission lines,
	whose width is drawn between $\sigma=$800 km/s and $\sigma=$2000 km/s 
	and with maximum heights between 
	5\% to 25\% of the continuum level. Finally, we add the same amount
	of noise as estimated for  the DEEP spectrum.
	
	\item One of us (RFC) then fitted the continuum  
	using a similar technique applied to the DEEP spectrum, without a priori knowledge of the true continuum.
	\gdb{An example is shown in figure \ref{fig:spectra_cont_test}}
	
	\item 
	Finally, we calculated the PDF of the \mlya\ transmitted flux \gdb{using the true and fitted}
	continua \gdb{for both the raw normalized fluxes and for the regulated fluxes.}
	The results are illustrated in figure
	\ref{fig:pdf_cont_test}.   
\end{itemize}
We find that the continuum fitting does increase the probability of 
the highest flux levels, which can be seen in the last bin of the flux of 
Fig. 
\ref{fig:pdf_cont_test}. Although the effect is very small, it could 
lead to a significant bias in the estimate of $\gamma$. However,
once we have applied the 95-percentile regulation, the difference 
between the true and the recovered distributions are negligible.

\section{The Effect of Contamination by Metal Lines and LLS}\label{appendix:contaminants}
\begin{figure*}
  \vskip -0.2in
     \centering{\epsfig{file=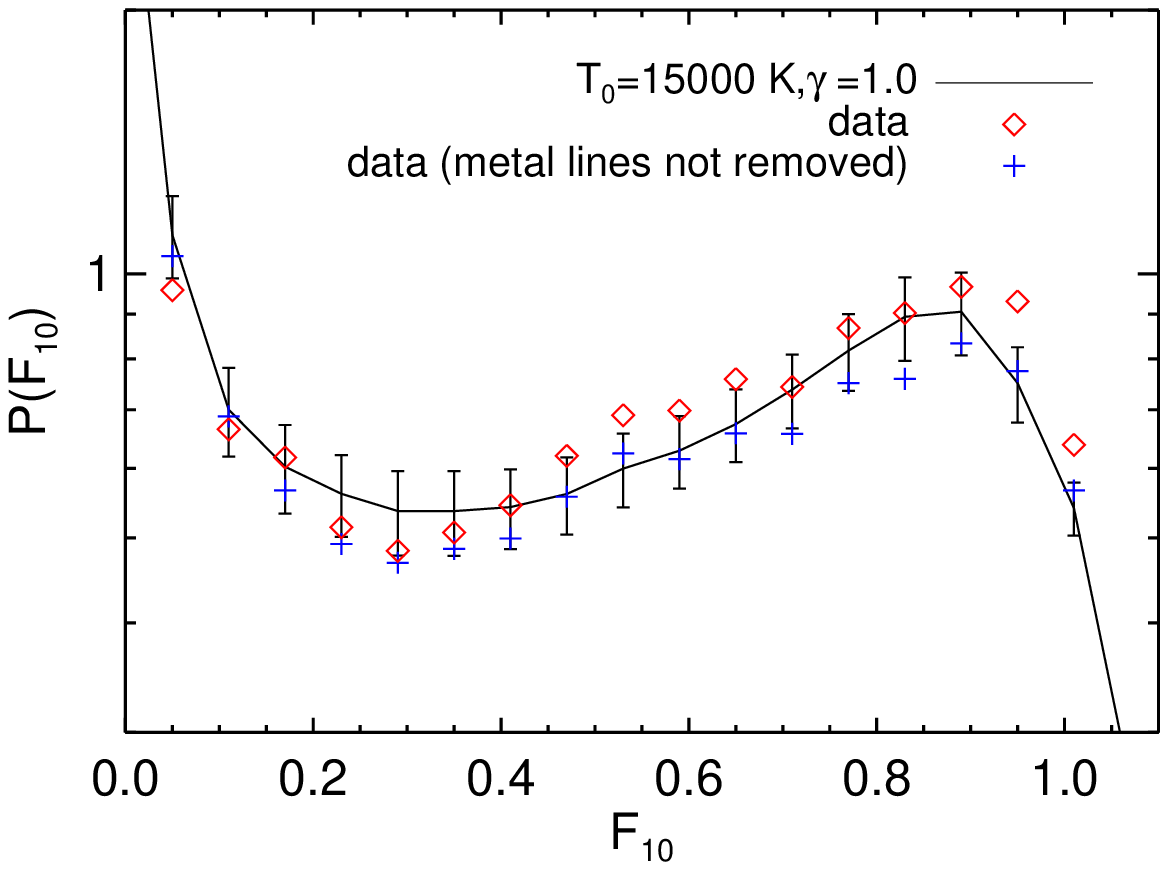,bb=21 -40 360 260 
       , width=0.36\textwidth,clip}
     \epsfig{file=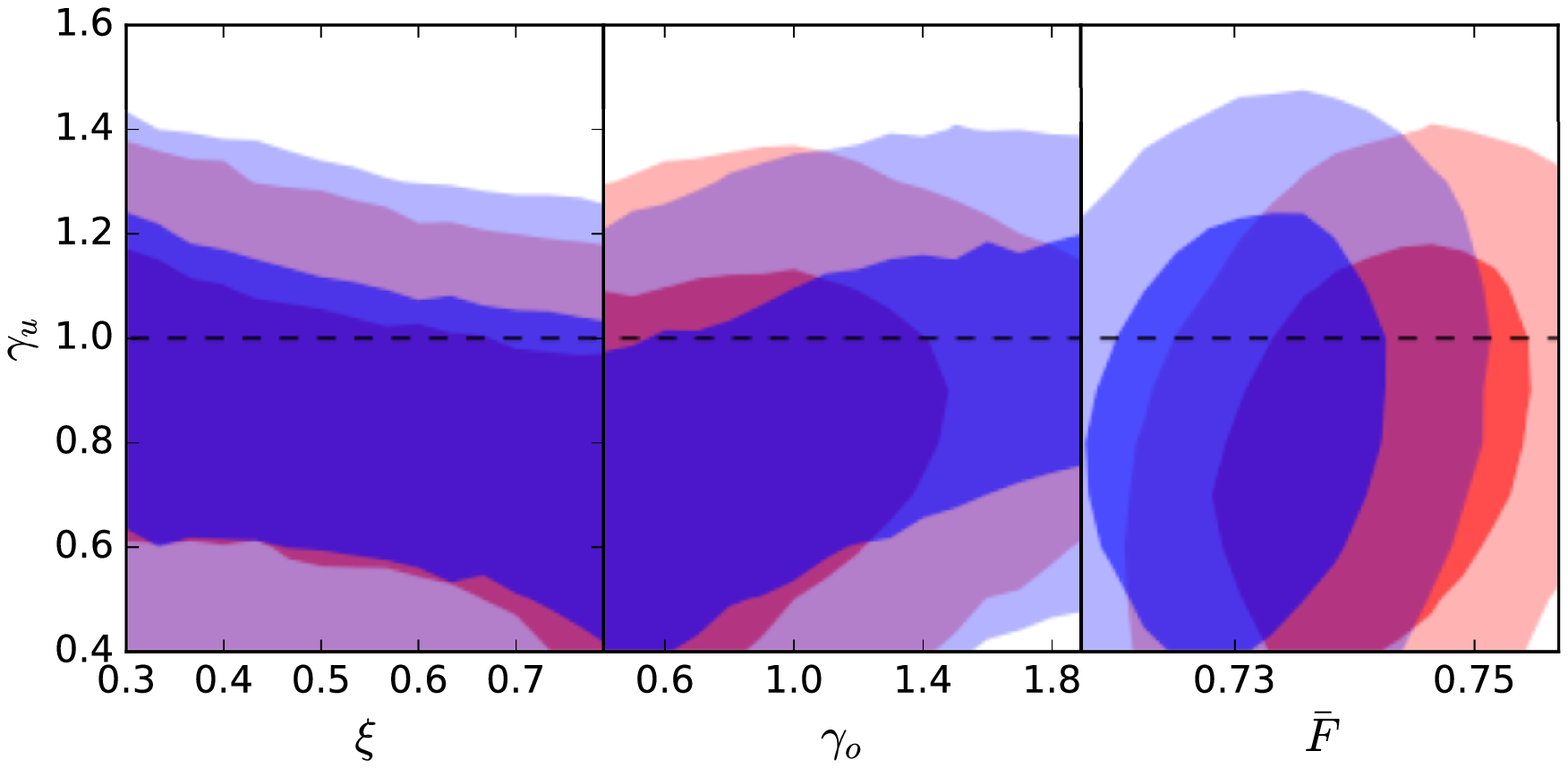,bb=20 -20 520 300
       , width=0.6\textwidth,clip}}
       
	\centering{}

   \vskip -0.25in
   \caption{\label{fig:contaminants}
   \emph{The effect of contamination from metal lines and LLS:} 
   on the left we plot the transformed and regulated PDF with metals
   and LLS identified and masked (default case, red diamonds) and 
   calculated from all the pixels in the \mlya\ forest (red crosses),
   before removing contaminants. The black solid lines represents an
   isothermal model with $T_0=15000$ K, $\xi=0.8$ and the mean flux
   set to $\bar{F}=0.7371$ (i.e. the observed value at z=$2.75$).
   In the right panel we plot the contours of the 68\% and 95\% 
   confidence levels in the parameter space of the broken TDR, 
   projected onto the $\xi$-$\gamma_u$,$\gamma_o$-$\gamma_u$ and 
   $\bar{F}$-$\gamma_u$ planes. The contours are calculated by 
   running the MCMC analysis on the data with metal masked (default,
   red) and on the whole, unmasked spectrum (blue). 
   }     
\end{figure*}
\begin{figure*}
\center
\includegraphics[width=\textwidth]{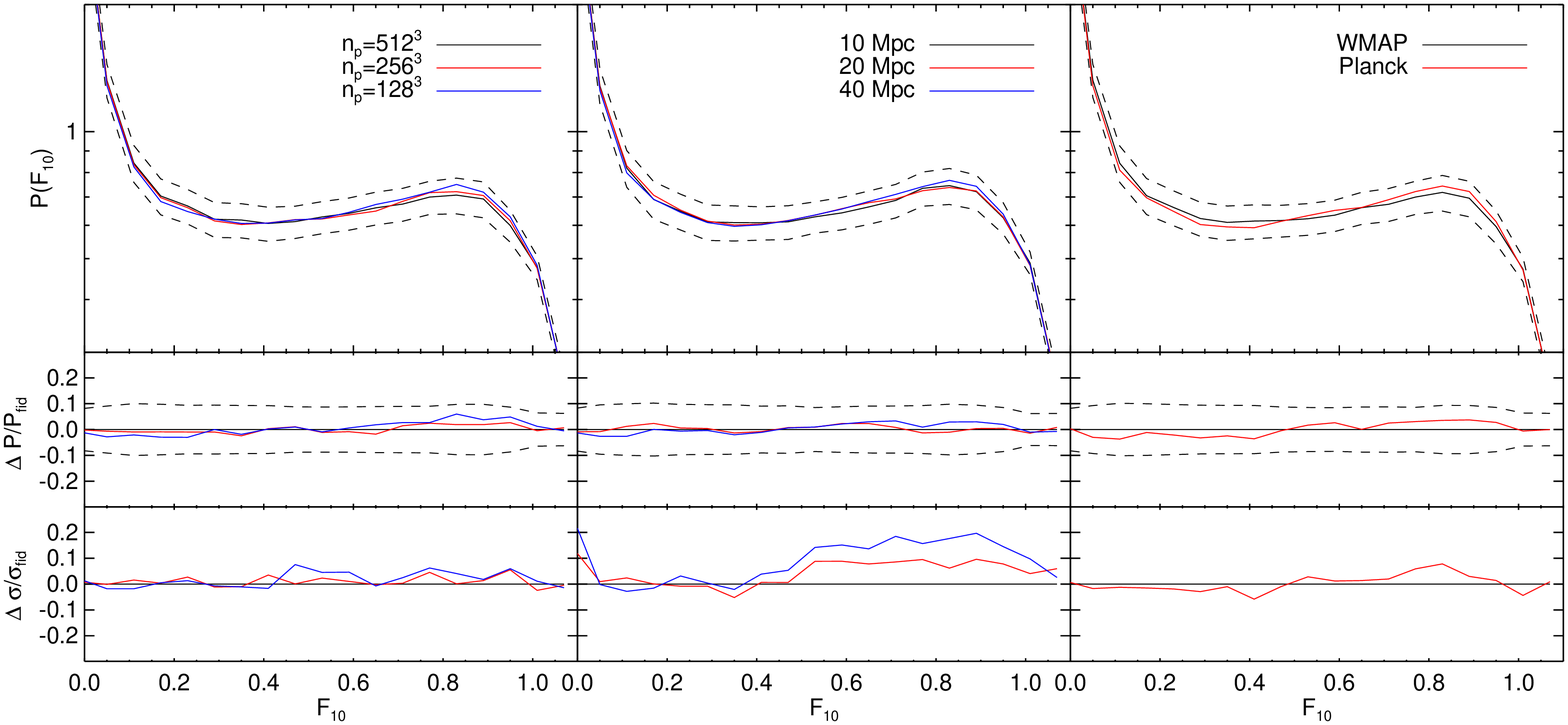}
\caption{\label{fig:resolution_convergence} Convergence test of the 
transformed and regulated PDF in our hydrodynamical simulations.
We show the effect of resolution (left), box size (middle) and 
cosmology (right). \review{The upper panels show the PDFs of the fiducial
runs and of the test runs, in the middle} panels we plot the relative deviation
of the PDF from the fiducial simulation, compared to the square root of the 
diagonal element of the covariance matrix (dashed black lines).
In the bottom panel we present instead the relative deviation of the 
diagonal errors. All the simulations used in this test have the 
smoothing parameter set to $\xi=1.45$.
\emph{Left panels:} results for  three simulations with box size 10 Mpc$/h$ with 
512$^3$ (fiducial,black), $256^3$ (red) 128$^3$ (blue) gas particles are shown. 
\review{The PDF and the diagonal errors are converged to a few-percent level.}
\emph{Central panels:} The fiducial model is a 10 Mpc$/h$ simulation
with 128$^3$ gas particles. We compare its PDF and the diagonal errors
to those calculated from  a 20 Mpc$/h$ simulation with 256$^3$ gas particles 
(in red), and a 40 Mpc$/h$ simulation with 512$^3$ particles 
\review{(in blue)}. 
While the mean PDF is converged to a small fraction of the estimated
uncertainty, the error themselves show a significant variation when the
box size is increased. \review{We correct the covariance matrix in our analysis 
to take this into account (see text).}
\emph{Right panels:} The black model uses the same cosmological
parameter we have used throughout this paper, while the red curves
are calculated assuming the best-fit values from the first data release
of Planck \citep{Planck}. \review{The differences are always significantly smaller 
than the statistical errors.}
}
\end{figure*}
The exceptional data quality of the DEEP spectrum facilitates 
the identification of narrow lines in the forest caused by 
metal absorption and of strong lines caused systems with a large
hydrogen column density (see \S~\ref{sec:contaminants} for a description
of our approach in removing metal lines and LLS). 
There is however an unavoidable uncertainty due to the possibility 
of blending of  the \mlya\ forest absorption with  these contaminants.

In this appendix we assess the impact of contamination by 
showing the difference in the PDF and in the results if the 
presence of metals and LLS were \emph{totally} ignored. 

In the left panel of Fig. \ref{fig:contaminants} we show the 
transformed and regulated PDF from the data as it was used in our
default analysis (red diamonds) and calculated from \emph{all 
the pixels} in the \mlya\ forest of the spectrum (blue crosses),
also those  that were flagged as contaminated. The PDF of an 
isothermal model is plotted in black for reference. We see 
indeed a slight change in the shape of the PDF, with in particular 
a decrease in the occurrence of high-flux pixels. 
This is a natural consequence of the  
increase in the average opacity when metal lines are included. 
The right panel of figure \ref{fig:contaminants} explicitly
reports the bias in the constraints we would get in the case 
of the broken TDR: the most relevant bias is the shift of the 
mean flux towards lower values in the contaminated case (blue  
contours) compared to the default simulation (red), and the broadening
of the constraints on the slope $\gamma_o$ in over-dense regions.
The posterior distribution of $\gamma_u$ and its degeneracy 
with the pressure smoothing parameter $\xi$ are only marginally affected,
demonstrating the robustness of our results with respect
to this possible source of bias. 

We argue that since metal lines and LLS have high optical depths,
they mainly alter the shape of the PDF at low values of $F$ (and
of course of $F_{10}$). The modification induced on the high-flux
part of the PDF is mainly attributable to the change of the overall 
mean flux which follows the removal of many \review{dark} pixels.

\section{Convergence tests}\label{app:convergence}

We verify the convergence  of the transformed flux PDF  with respect 
to  resolution and  box size of the simulations. We use as reference a simulation
with box size of 10 Mpc$/h$ and $512^3$ gas particles. We then compare the transformed
PDF of this default run with those calculated  from
two simulations with the same box size and $256^3$
and $128^3$ gas particles, respectively. All the simulations  have been performed with  a pressure smoothing parameter of $\xi=1.45$,
and for this test we use the actual temperature distribution
of the hydrodynamical simulation, instead of the post-processed temperature-density
relation. The parameters obtained by fitting the phase space 
are $T_0=14000$ K and $\gamma=1.54$ at $z=3$.
We then calculate the PDF of the transformed flux $F_{10}$,
after applying the continuum regulation. 
We also calculate the covariance matrix of the same statistic. 
The result of this test is shown in the left panel of figure 
\ref{fig:resolution_convergence}. The PDF is converged within the 
estimated errorbars at all flux levels. The errors, taken as the 
square root of the diagonal elements of the covariance matrix, are also reasonably converged,
with a difference of at most  10\%  between the $256^3$ and the $512^3$ simulations. 

An analogous test is carried out in order to verify the convergence with respect
to the box size of the simulations. We use three simulations with box size 10,20 and 40 Mpc$/h$ with
$128^3,256^3$ and $512^3$ gas particles respectively.  With this choice 
 the three simulations have the same resolution, so that we can
isolate the effect of volume sampling alone. The central panel of 
figure \ref{fig:resolution_convergence}
shows that the PDF is converged at the few percent level, significantly
better than the estimated statistical uncertainty. The standard deviations are 
less well converged than in the resolution test, revealing the effect of cosmic variance
in the simulations with larger box size. We find that \review{variations 
could be as large as as 20\%.
We therefore correct the standard deviations in our simulations  
by a factor $\iota=1.2$, which corresponds to a correction to variances of $\iota^2=1.44$. For this reason, we multiply each element
of the covariance matrix employed in the likelihood in formula
\ref{eq:likelihood} by 1.44. }

\begin{figure}
\begin{center}
\includegraphics[trim={1cm 0 1cm 0},clip,width=\columnwidth]{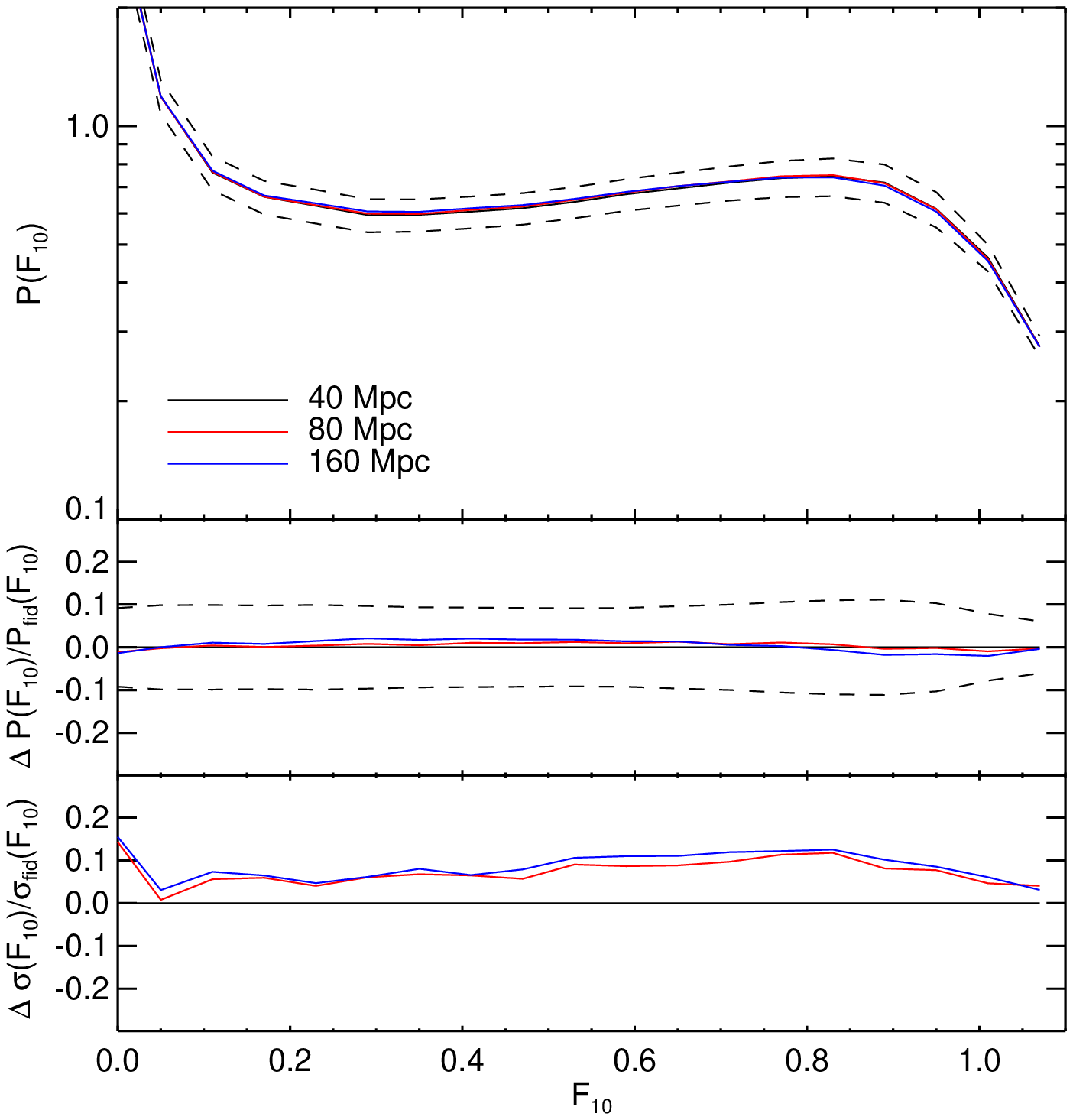}
\caption{\label{fig:prace_convergence} 
\review{Same as the central panel of Fig.~\ref{fig:resolution_convergence}, 
but considering} three 
runs from the Sherwood set of simulations \protect\citep{Sherwood}. 
The boxes are  40,80 and 160 Mpc$/h$ of size with $2\times,512^3,
1024^3,2048^3$ particles respectively, which \review{ matches the 
resolutions of the simulation shown in Fig.~\ref{fig:resolution_convergence}.}
Compared to the previous test on smaller boxes,
the convergence of the errors is significantly improved, 
down to a level of 5\% at most
for the 80 Mpc$/h$ box (we remind the reader that the $F_{10}=0$ bin is
excluded from the analyses). 
We reckon that this level of convergence is satisfactory for the degree of precision of this work, 
and a slight change in our covariance estimate would not qualitatively
modify any of our conclusions.
}

\end{center}
\end{figure}
Since error estimation of the flux PDF is known to be a delicate process in the flux
PDF \citep{Rollinde2013} we extend our convergence test to bigger boxes 
taking advantage of a new set of high-resolution hydrodynamical runs,
the Sherwood simulations \citep{Sherwood}. 
These simulations were run using the Gadget-3 code \citep{Gadget2}, 
assuming the  UV background and photoheating rates from the model of \cite{Haardt12}. More details can be found in Bolton \emph{et al.} (submitted)
and \cite{Keating2016}.
We used three simulations with size  40,80 and 160  Mpc$/h$ and $512^3,1024^3,2048^3$
gas particles, respectively. The resolution of these runs is the same as that of
 simulations used for the previous convergence test, although 
the cosmology and the assumption on the thermal history are slightly
different. 
The results are shown in figure~\ref{fig:prace_convergence}. The convergence
achieved at 40 Mpc$/h$ is \review{excellent with respect to the PDF, and 
better than 10\% for the  estimated errors. This should be 
sufficient for our purposes.}

\section{Dependence on Cosmological Parameters}
We also test the dependency of the flux PDF of  
cosmological parameters, by comparing our default simulations, which assumes the 
best-fit cosmological parameters of the WMAP data, to a simulated box run with the
cosmological parameters as constrained from the first data release of 
Planck \citep{Planck}:
$\Omega_m=0.274,\Omega_b=0.0457,n_s=0.968,H_0=70.2$ km s$^{-1}$
Mpc$^{-1}$ and $\sigma_8=0.816$.
The result is shown in the rightmost panel of figure
\ref{fig:resolution_convergence}: the variation to the change in 
cosmology is always less significant than the estimated errors. 

\review{\section{Tabulated Values of the PDF}
\begin{center}
\begin{table}
\centering
\begin{tabular}{ccc}
\hline\hline\\
$\langle F_{10}^{(a)} \rangle$        & $P(F_{10})^{(b)}$ & $\Delta P(F_{10})^{(c)}$  \\
\hline\\
-0.01   & 4.205 & 0.410  \\
0.05		& 0.958 & 0.119  \\
0.11		& 0.665 & 0.081  \\
0.17		& 0.617 & 0.070  \\
0.23		& 0.514 & 0.060  \\
0.29	 	& 0.483 & 0.059  \\
0.35 	& 0.507 & 0.060  \\
0.41 	& 0.545 & 0.057  \\
0.47 	& 0.620 & 0.057  \\
0.53 	& 0.690 & 0.058  \\
0.59 	& 0.699 & 0.060  \\
0.65 	& 0.759 & 0.064  \\
0.71 	& 0.743 & 0.071  \\
0.77 	& 0.868 & 0.082  \\
0.83 	& 0.902 & 0.097  \\
0.89 	& 0.967 & 0.098  \\
0.95 	& 0.930 & 0.074  \\
1.01 	& 0.638 & 0.038  \\
1.07 	& 0.239 & 0.019  \\
\hline\hline
\end{tabular}
\caption{\review{Tabulated values of the PDF of the regulated and 
transformed flux.
(a) Central value of the flux bin.
(b) Probability density of $F_{10}$.
(c) Error on the PDF estimated from a fiducial simulation (see text).} }
\label{tab:pdf}
\end{table}
\end{center}
We report in Table~\ref{tab:pdf} the values of the PDF of the 
regulated and transformed flux from HE0940-1050. We emphasize that for a meaningful
comparison all data properties should be included in the models (see
\S~\ref{sec:forward_modelling}). In order to add noise to the synthetic spectra,
we suggest to add Gaussian fluctuations to each pixel (which should be 2.5 km/s
wide), using the amplitude
given by the flux-dependent expression $\sigma (F)=\sqrt{\sigma_0^2+F(\sigma_c^2-
\sigma_0^2)}$, with $\sigma_0=0.0028$ and $\sigma_c=0.0088$.
We have checked that this formula represents an excellent approximation of the 
more comprehensive treatment described in \S~\ref{sec:forward_modelling}.
Continuum regulation and optical-depth
rescaling procedures must be applied to simulated spectra consistently 
with what we have done in this paper (\S~\ref{sec:continuum_regulation},~\ref{sec:transformation}). 
Finally, note that the given errors are 
estimated from a fiducial simulation with $T_0=15000$ K and $\gamma=1$ and 
should be used only for illustrative purposes. A rigorous statistical comparison
 with the PDF data  requires the prediction of the full covariance matrix
from the model, as explained in \S~\ref{sec:error_estimation}.}

\bsp	
\label{lastpage}
\end{document}